\if@article
\def\tableofcontents{\section*{\contentsname
\@mkboth{\contentsname}{\contentsname}}
\@starttoc{toc}}

\def\listoffigures{\section*{\listfigurename\@mkboth
{\listfigurename}{\listfigurename}}\@starttoc{lof}}
\def\listoftables{\section*{\listtablename\@mkboth
{\listtablename}{\listtablename}}\@starttoc{lot}}

\def\thebibliography#1{\section*{\refname\@mkboth
 {\refname}{\refname}}\list
 {[\arabic{enumi}]}{\settowidth\labelwidth{[#1]}\leftmargin\labelwidth
 \advance\leftmargin\labelsep
 \usecounter{enumi}}
 \def\newblock{\hskip .11em plus .33em minus -.07em}
 \sloppy
 \sfcode`\.=1000\relax}

\def\theindex{\@restonecoltrue\if@twocolumn\@restonecolfalse\fi
\columnseprule \z@
\columnsep 35pt\twocolumn[\section*{\indexname}]
\@mkboth{\indexname}{\indexname}\thispagestyle{plain}\parindent\z@
\parskip\z@ plus .3pt\relax\let\item\@idxitem}

\def\@part[#1]#2{\ifnum \c@secnumdepth >\m@ne \refstepcounter{part}
\addcontentsline{toc}{part}{\thepart \hspace{1em}#1}\else
\addcontentsline{toc}{part}{#1}\fi { \parindent 0pt \raggedright
\ifnum \c@secnumdepth >\m@ne \Large \bf \partname\ \thepart \par
\nobreak \fi \huge
\bf #2\@mkboth{}{}\par } \nobreak \vskip 3ex \@afterheading }

\if@twoside \def\ps@headings{\def\@oddfoot{}\def\@evenfoot{}\def\@evenhead{\rm
\thepage\hfil \sl \leftmark}\def\@oddhead{\hbox{}\sl \rightmark \hfil
\rm\thepage}\def\sectionmark##1{\@mkboth {\ifnum \c@secnumdepth
>\z@
 \thesection\hskip 1em\relax \fi ##1}{}}\def\subsectionmark##1{\markright
{\ifnum \c@secnumdepth >\@ne
 \thesubsection\hskip 1em\relax \fi ##1}}}
\else \def\ps@headings{\def\@oddfoot{}\def\@evenfoot{}\def\@oddhead{\hbox
{}\sl \rightmark \hfil \rm\thepage}\def\sectionmark##1{\markright
{\ifnum \c@secnumdepth >\z@
 \thesection\hskip 1em\relax \fi ##1}}}
\fi
\def\ps@myheadings{\def\@oddhead{\hbox{}\sl\rightmark \hfil
\rm\thepage}\def\@oddfoot{}\def\@evenhead{\rm \thepage\hfil\sl\leftmark\hbox
{}}\def\@evenfoot{}\def\sectionmark##1{}\def\subsectionmark##1{}}

\if@titlepag
\def\abstract{\titlepage
\null\vfil
\begin{center}
{\bf \abstractname}
\end{center}}
\def\endabstract{\par\vfil\null\endtitlepage}
\else
\def\abstract{\if@twocolumn
\section*{\abstractname}
\else \small
\begin{center}
{\bf \abstractname\vspace{-.5em}\vspace{0pt}}
\end{center}
\quotation
\fi}
\def\endabstract{\if@twocolumn\else\endquotation\fi}
\fi
\fi

\def\refname{References}

\def\contentsname{Contents}
\def\listfigurename{List of Figures}
\def\listtablename{List of Tables}
\def\indexname{Index}
\def\figurename{Figure}
\def\tablename{Table}
\def\partname{Part}
\def\abstractname{Abstract}

\def\fnum@figure{\figurename\ \thefigure}
\def\fnum@table{\tablename\ \thetable}

\if@twoside \def\ps@headings{%
\def\@oddfoot{}
\def\@evenfoot{}
\def\@evenhead{\underline{\hbox to \textwidth{\strut%
               \rm\thepage\hfil \sl \leftmark}}}
\def\@oddhead{\underline{\hbox to \textwidth{\strut%
              \hbox{}\sl \rightmark \hfil \rm\thepage}}}
        \def\sectionmark##1{\markboth {\uppercase{\ifnum \c@secnumdepth
>\z@
 \thesection\hskip 1em\relax \fi ##1}}{}}\def\subsectionmark##1{\markright
{\ifnum \c@secnumdepth >\@ne
 \thesubsection\hskip 1em\relax \fi ##1}}}
\else \def\ps@headings{%
\def\@oddfoot{}
\def\@evenfoot{}
\def\@oddhead{\underline{\hbox to \textwidth{\strut%
              \hbox{}\sl \rightmark \hfil \rm\thepage}}}
\def\sectionmark##1{\markright
{\uppercase{\ifnum \c@secnumdepth >\z@
 \thesection\hskip 1em\relax \fi ##1}}}}
\fi
\def\ps@myheadings{%
\def\@oddhead{\underline{\hbox to \textwidth{\strut%
              \hbox{}\sl\rightmark \hfil \rm\thepage}}}
\def\@oddfoot{}
\def\@evenhead{\underline{\hbox to \textwidth{\strut%
               \rm \thepage\hfil\sl\leftmark\hbox{}}}}
\def\@evenfoot{}
\def\sectionmark##1{}\def\subsectionmark##1{}}

\def\@xfloat#1[#2]{\ifhmode \@bsphack\@floatpenalty -\@Mii\else
   \@floatpenalty-\@Miii\fi\def\@captype{#1}\ifinner
      \@parmoderr\@floatpenalty\z@
    \else\@next\@currbox\@freelist{\@tempcnta\csname ftype@#1\endcsname
       \multiply\@tempcnta\@xxxii\advance\@tempcnta\sixt@@n
       \@tfor \@tempa :=#2\do
                        {\if\@tempa h\advance\@tempcnta \@ne\fi
                         \if\@tempa t\advance\@tempcnta \tw@\fi
                         \if\@tempa b\advance\@tempcnta 4\relax\fi
                         \if\@tempa p\advance\@tempcnta 8\relax\fi
         }\global\count\@currbox\@tempcnta}\@fltovf\fi
    \global\setbox\@currbox\vbox\bgroup
    \def\baselinestretch{1}\small\normalsize
    \boxmaxdepth\z@
    \hsize\columnwidth \@parboxrestore}
\long\def\@footnotetext#1{\insert\footins{\def\baselinestretch{1}\footnotesize
    \interlinepenalty\interfootnotelinepenalty
    \splittopskip\footnotesep
    \splitmaxdepth \dp\strutbox \floatingpenalty \@MM
    \hsize\columnwidth \@parboxrestore
   \edef\@currentlabel{\csname p@footnote\endcsname\@thefnmark}\@makefntext
    {\rule{\z@}{\footnotesep}\ignorespaces
      #1\strut}}}

\def\singlespace{%
\vskip\parskip%
\vskip\baselineskip%
\def\baselinestretch{1}%
\ifx\@currsize\normalsize\@normalsize\else\@currsize\fi%
\vskip-\parskip%
\vskip-\baselineskip%
}

 \documentstyle[12pt,twoside,TKdiss]{article}
 \def\3{\ss}
 \renewcommand{\thesection}{\arabic{section}}
 \renewcommand{\thesubsection}{\thesection.\arabic{subsection}}
 \renewcommand{\theequation}{{\protect\thesection.\arabic{equation}}}
 \renewcommand{\thetable}{{\protect{\bf\thesection.\arabic{table}}}}
 \renewcommand{\thefigure}{{\protect\bf\thesection.\arabic{figure}}}
 \newcounter{abc}
 \newcounter{fnmarkone}
 
 \newcounter{FIG}
 \newcounter{TABLE}
 \newcounter{FigBlockBose}
 \newcounter{SecBlockBose}

 \newcounter{FigPFlattice}
 \newcounter{SecPFlattice}

 \newcounter{FigAbose}
 \newcounter{SecAkernel}

 \newcounter{appSF}
 \renewcommand{\theappSF}{\Alph{appSF}}
 \newcounter{appKernels}
 \renewcommand{\theappKernels}{\Alph{appKernels}}
 
 \renewcommand{\figurename}{{\protect\bf{Fig.}}}
 \renewcommand{\tablename}{{\protect\bf{Table}}}
 \newcommand{\Section}[1]{\par\cleardoublepage
                          \mbox{}\par\mbox{}\vspace{2.5 cm}\mbox{}\par
                          \mbox{}
                          \begin{center}\section{#1}\end{center}
                          \setcounter{equation}{0}
                          \setcounter{figure}{0}
                          \setcounter{table}{0}}
 
 \newcommand{\Sectionstar}[1]{\par\cleardoublepage
                          \mbox{}\par\mbox{}\vspace{2.5 cm}\mbox{}\par
                          \mbox{}
                              \begin{center}\section*{#1}\end{center}
                          \setcounter{equation}{0}
                          \setcounter{figure}{0}
                          \setcounter{table}{0}}

 \newcommand{\Subsection}[1]{\begin{center}\subsection{#1}\end{center}}
 \newcommand{\Subsectiontocentry}[2]{
                                     \begin{center}
                                     \subsection[#1]{#2}\end{center}}
 \newcommand{\SubSubsection}[1]{%
                           \begin{center}\subsubsection{#1}\end{center}}
 
 \newcommand{\Subsubsection}[1]{\par\smallskip{\small\bf\noindent #1.}
                                \newline\indent}
 \newcommand{\Figure}[2]{\begin{figure}[tbp]{\vspace*{#1}}
                                     \caption{{\protect\small{\em #2}}}
                         \end{figure}}
 
\def\thebibliography#1{\Sectionstar{References\markboth
 {\sc References}{}}\list
 {[\arabic{enumi}]}{\settowidth\labelwidth{[#1]}\leftmargin\labelwidth
 \advance\leftmargin\labelsep
 \usecounter{enumi}}
 \def\newblock{\hskip .11em plus .33em minus -.07em}
 \sloppy
 \sfcode`\.=1000\relax}

 \def\bbbn{{\rm I\!N}} 
 \def\bbbone{{\mathchoice {\rm 1\mskip-4mu l} {\rm 1\mskip-4mu l}
 {\rm 1\mskip-4.5mu l} {\rm 1\mskip-5mu l}}}
 \def\bbbc{{\mathchoice {\setbox0=\hbox{$\displaystyle\rm C$}\hbox{\hbox
 to0pt{\kern0.4\wd0\vrule height0.9\ht0\hss}\box0}}
 {\setbox0=\hbox{$\textstyle\rm C$}\hbox{\hbox
 to0pt{\kern0.4\wd0\vrule height0.9\ht0\hss}\box0}}
 {\setbox0=\hbox{$\scriptstyle\rm C$}\hbox{\hbox
 to0pt{\kern0.4\wd0\vrule height0.9\ht0\hss}\box0}}
 {\setbox0=\hbox{$\scriptscriptstyle\rm C$}\hbox{\hbox
 to0pt{\kern0.4\wd0\vrule height0.9\ht0\hss}\box0}}}}
 \def\bbbz{{\mathchoice {\hbox{$\sf\textstyle Z\kern-0.4em Z$}}
 {\hbox{$\sf\textstyle Z\kern-0.4em Z$}}
 {\hbox{$\sf\scriptstyle Z\kern-0.3em Z$}}
 {\hbox{$\sf\scriptscriptstyle Z\kern-0.2em Z$}}}}
 \renewcommand{\baselinestretch}{1.1}
 
 \renewcommand{\d}{\displaystyle}
 \renewcommand{\hat}[1]{\widehat{#1}}
 \renewcommand{\L}{L_b}
 \renewcommand{\tilde}[1]{\widetilde{#1}}
 \renewcommand{\H}{{\cal H}}
 \newcommand{\NextFig}{\setcounter{FIG}{\value{figure}}%
                       \addtocounter{FIG}{1}\thesection.\theFIG}

 \newcommand{\PreviousFig}{\setcounter{FIG}{\value{figure}}%
                           \thesection.\theFIG}

 \newcommand{\subs}[1]{\mbox{$\!$\protect\scriptsize\it#1}}
 \newcommand{\av}[1]{\raisebox{-7pt}{$\stackrel{\mbox{av}}%
                                       {\scriptstyle #1}$}}
 \newcommand{\bc}{b.~c.\ }
 
 \newcommand{\bpl}{\mbox{{\large\bf(}}}
 \newcommand{\bpr}{\mbox{{\large\bf)}}}

 \newcommand{\emu}{e_{\mu}}
 \newcommand{\en}{e^{(n)}}
 
 \newcommand{\equ}[1]{(\ref{#1})}
 \newcommand{\esk}{\enspace ,}
 \newcommand{\esp}{\enspace .}
 \newcommand{\etamu}{\eta_{\mu}}

 \newcommand{\lambdanull}{\lambda_{0}(x)}

 \newcommand{\lsim}{\raisebox{-3pt}{$\stackrel{<}{\sim}$}}
 \newcommand{\mcr}{m_{\subs{cr}}^2}

 \newcommand{\normC}{\| C(x,z) \|}
 \newcommand{\ol}{\overline}
 \newcommand{\oopt}{\omega_{\subs{opt}}}
 \newcommand{\phin}{\phi^{(n)}}
 \newcommand{\phinpo}{\phi^{(n+1)}}
 \newcommand{\phint}{\tilde{\phi}^{(n)}}
 \newcommand{\re}{\mbox{\rm Re\,}}
 \newcommand{\rhs}{r.\,h.\,s.\ }
 \newcommand{\rGS}{\rho_{\subs{GS}}}
 \newcommand{\rJ}{\rho_{\subs J}}
 \newcommand{\rn}{r^{(n)}}
 \newcommand{\rom}{\rho_{\omega}}
 \newcommand{\ropt}{\rho_{\oopt}}

 \newcommand{\txah}{{\textstyle\frac{a}{2}}}
 
 \newcommand{\txeh}{{\textstyle\frac{1}{2}}}

 \newcommand{\vpn}{\varphi^{(n)}}
 \newcommand{\zhdh}{2^{d/2}}
 \newcommand{\A}{{\cal A}}
 \newcommand{\Astar}{\A^{\ast}}
 
 \newcommand{\C}{{\cal C}}
 \newcommand{\Cstar}{C^{\ast}}
 
 \newcommand{\D}{{\cal D}}
 
 \newcommand{\Dm}{\triangle m^2}
 \newcommand{\Dirac}{\mbox{$\not\!\!D$}}
 
 \newcommand{\Gaw}{Gaw\c{e}dzki\ }
 \newcommand{\Lah}{\Lambda_{a/2}}

 \newcommand{\LAPD}{\Delta_{D,x}}
 \newcommand{\LAPN}{\Delta_{N,x}}
 
 \newcommand{\Lnull}{\Lambda^0}
 \newcommand{\Lone}{\Lambda^1}
 \newcommand{\Ltwo}{\Lambda^2}
 \newcommand{\LN}{\Lambda^N}
 \newcommand{\M}{{\cal M}}
 \newcommand{\Nc}{N_{\subs{c}}}
 
 \newcommand{\Tr}{\mbox{\rm Tr\,}}

\begin{document}
 \pagestyle{empty}
 \pagestyle{plain}
 \pagenumbering{roman}

 \noindent{\tt DESY $92-158$ \hfill ISSN $0418-9833$}\\
 {\tt November 1992}
 \renewcommand{\thefootnote}{{\protect\fnsymbol{footnote}}}

 \begin{center}
 \vspace{4cm}

 {\LARGE  Multigrid Methods for the Computation of} \\
 \medskip
 {\LARGE  Propagators in Gauge Fields
          \footnote{Work supported by Deutsche Forschungsgemeinschaft.}}

 \addtocounter{footnote}{5}
 \vspace{3cm}
         Thomas Kalkreuter \footnote{E-mail: I02KAL@DSYIBM.DESY.DE} \\
         \smallskip
    {\em II. Institut f\"ur Theoretische Physik
         der Universit\"at Hamburg, \\
         Luruper Chaussee 149,
         W-2000 Hamburg 50, Germany}
 \end{center}

 \vspace{1.5cm}\mbox{}

 \vfill

 \vfill\mbox{}
 \newpage

 \mbox{}\vfill
 \renewcommand{\abstractname}{Abstract}
 \begin{abstract}
 \noindent
 Multigrid methods were invented for the solution of discretized
 partial differential equations in order to overcome the slowness
 of traditional algorithms by updates on various length scales.
 In the present work generalizations of multigrid methods for
 propagators in gauge fields are investigated.
 Gauge fields are incorporated in algorithms in a covariant way.
 This avoids the necessity for gauge fixing in computations of
 propagators.
 The kernel $C$ of the restriction operator which averages from
 one grid to the next coarser grid is defined by projection
 on the ground-state of a local Hamiltonian
 (e.\ g.\ a block-local approximation of the fermion matrix).
 The idea behind this definition is that the appropriate notion of
 smoothness depends on the dynamics.
 In traditional algorithms the lowest mode of the Hamiltonian is
 responsible for critical slowing down, and this mode should be
 represented as well as possible on coarser grids.
 The ground-state projection choice of $C$ is usable in arbitrary
 space-time dimension $d$ and for arbitrary gauge group.
 We discuss proper averaging operations for bosons and for staggered
 fermions.
 An efficient algorithm for computing $C$ numerically is presented.
 The averaging kernels $C$ can be used not only in deterministic
 multigrid computations, but also in multigrid Monte Carlo simulations,
 and for the definition of block spins and blocked gauge fields in
 Monte Carlo renormalization group studies of gauge theories.
 Actual numerical computations of kernels and propagators are
 performed in compact four-dimensional $SU(2)$ gauge fields.
 We prove that our proposals for block spins are ``good'',
 using renormalization group arguments.
 The argumentation uses an ``optimal'' interpolation kernel~$\A$
 which is associated with~$C$.
 This $\A$ is a gauge covariant generalization of a kernel which was
 used successfully in rigorous works on constructive quantum field
 theory by \Gaw and Kupiainen.
 A central result of the present work is that {\em the multigrid method
 works in arbitrarily disordered gauge fields, in principle\/}.
 It is proved that computations of propagators in gauge fields
 without critical slowing down are possible when one uses the ideal
 $\A$.
 Unfortunately, the idealized algorithm is not practical, but it was
 important to answer questions of principle.
 Practical multigrid methods are able to outperform the conjugate
 gradient algorithm in case of bosons.
 The case of staggered fermions is harder.
 Multigrid methods give considerable speed-ups compared to
 conventional relaxation algorithms, but on lattices up to $18^4$
 conjugate gradient is superior.
 A modification which can be thought of as ``updating on a layer
 consisting of a single site'' leads to an improvement of relaxation
 algorithms in case of bosons.
 However, we feel unable to predict whether the analogous method
 pays for staggered fermions on lattices of realizable sizes, because a
 volume effect remains with respect to how long it takes until
 errors decay exponentially.
 \end{abstract}
 \vfill\mbox{}
 \renewcommand{\thefootnote}{\arabic{footnote})}
 \setcounter{footnote}{0}
 \newpage
 \tableofcontents
 \newpage
 \mbox{}
 \newpage
 \pagestyle{myheadings}
 \pagenumbering{arabic}

 \Section{Introduction}
 \markboth{{\rm\thesection.}\ {\sc Introduction}}{}
 The lattice regularization provides presently the only known
 nonperturbative tool to investigate quantum field theories
 quantitatively.
 In the context of quantum chromodynamics (QCD) it was introduced by
 Wilson in 1974~\cite{Wil}.

 \smallskip
 Lattice calculations are performed in Euclidean space.
 Minkowski space field theories \cite{ItzZub,Ryd} are transformed into
 Euclidean theories \cite{Sym,OstSch,OstSei,Sei,GliJaf}
 by analytic continuation to imaginary time.
 Through this procedure field theory is mapped onto a system of
 classical statistical mechanics \cite{Hua}.
 This mapping becomes manifest in the path integral representation of
 Green's functions; see e.\ g.\ Refs.~\cite{Jer,Roe} for reviews.

 \smallskip
 It is desirable to carry through analytic calculations on the lattice.
 Maybe the most prominent and important result in recent time is the
 solution of $\phi^4_4$ theory by L\"uscher and Weisz \cite{LueWei}
 which yielded a nonperturbative upper bound for the mass of the Higgs
 particle.

 \smallskip
 Besides analytic calculations the numerical investigation of
 lattice field theories is another important branch.
 Euclidean field theories can be simulated numerically by means of
 Monte Carlo methods \cite{HamHan,Bin,CreJacReb,Hee}.
 A collection of  ``early-days-papers'' is contained in Rebbi's
 book~\cite{Reb}.
 Further introductory references are \cite{KogSpin,KogQCD,CreBook,Rot}.

 \medskip
 In numerical simulations of lattice field theories one faces the
 problem of critical slowing down (CSD).
 Possible continuum limits can only be taken at those points in the
 phase diagram where phase transitions of second order arise.
 Expectation values of observables are determined through generating
 a Marcov chain in a Monte Carlo process.
 Subsequent field configurations in the Marcov chain are correlated.
 The correlation time determines, roughly speaking, the number of
 Monte Carlo sweeps which are necessary to obtain a statistically
 independent configuration.
 The problem of CSD consists of the phenomenon that correlation times
 diverge as critical points are approached.

 \smallskip
 Numerous algorithms have been tested and new ones developed in order to
 circumvent, or at least to mitigate, the problem of CSD.
 A very successful class of algorithms for spin models is the class
 of cluster algorithms.
 The first cluster simulation was performed by Swendsen and Wang
 \cite{SweWan} for the Ising model \cite{Isi,Hua} in two and three
 dimensions.
 The generalization of their algorithm to $O(N)$ models is due to
 Wolff \cite{Wol} and Hasenbusch \cite{Has}.
 Reviews of cluster algorithms can be found in~\cite{ClusterReviews}.

 \smallskip
 Other candidates for fighting CSD are multigrid (MG) algorithms.
 Their basis is the following observation.
 The slowness of conventional algorithms is due to the fact that
 field configurations are updated only locally.
 The slow modes are the long-wavelength modes
 (i.\ e.\ smooth modes in ordered systems).
 These modes, however, can be treated more efficiently on coarser
 lattices.
 Measured in units of lattice spacings, long wavelengths are effectively
 shortened on coarser scales.
 Parisi~\cite{Par} was the first one who pointed out that MG Monte
 Carlo methods would be promising candidates for tackling CSD.

 \smallskip
 MG proposals were made by H.\,Meyer-Ortmanns \cite{HMO}, and by
 Goodman and Sokal \cite{GooSok}.
 The latter authors presented the first MG Monte Carlo
 simulation of two-dimensional $\phi^4$ theory.
 Their approach was to generalize existing deterministic methods
 to stochastic methods.
 Simulations in their spirit were performed in
 Refs.~\cite{HSVGSP,HSVXY,GooSokSeries,LauVin}.

 \smallskip
 Mack~\cite{MacCargese} presented an MG approach to quantum field theory
 which is inspired by rigorous work in constructive field theory.
 This approach combines Monte Carlo simulations with analytical tools,
 namely with ideas from the renormalization group
 \cite{WilKog,Ami,Binetal} and with the theory of polymer systems
 \cite{GruKun,MacRGandME,MacPor,MacCargese,Sei,GliJaf}.

 \smallskip
 Mack's MG proposal has the additional advantage that it yields the
 effective action (in the sense of the renormalization group) and its
 derivatives for free~\cite{MacCargese,MacMey}.
 Moreover, one can extract infinite volume results from a simulation
 in a finite volume as shown by Palma~\cite{Pal} for pure
 $\phi^4$-theory.
 Palma found perfect agreement with the calculations of L\"uscher
 and Weisz \cite{LueWei}.
 Furthermore, it was pointed out in Ref.~\cite{MacCargese} that a
 reformulation of a theory as a polymer system on an MG offers
 the possibility of performing simulations for continuum systems
 without imposing a UV-cutoff.

 \smallskip
 A central point in Mack's MG philosophy is the need to minimize
 couplings between low and high frequency fields in order to map
 the original (nearly) critical system onto a noncritical system on
 the MG\@.
 As a consequence interpolation kernels should be smooth in order to
 fight CSD\@.
 The correctness of this contention was demonstrated in an impressive
 way by Hasenbusch and Meyer~\cite{HasMeyMac,HasMey}.

 \smallskip
 The use of ideas from the theory of polymer systems in conventional
 Monte Carlo simulations was proposed by Mack and Pinn~\cite{MacPin}.
 The present author~\cite{KalDiplom} showed how their proposal can be
 used for more efficient simulations of theories with improved lattice
 actions~\cite{SymImprovedAction}.
 Furthermore, the polymer approach can be viewed as a more general
 framework for cluster algorithms.
 A problem which one encounters in polymer algorithms is the
 possibility of a non-positive-definiteness of probability densities.
 The present author~\cite{KalNP1} made a proposal how to deal with
 systems with complex actions.

 \smallskip
 The problem of complex actions is also encountered in QCD
 simulations~\cite{Mon}, and this brings us to the problems related
 with lattice fermions.
 Fermionic degrees of freedom cannot be simulated straightforwardly
 because they are represented by Grassmann variables.
 The common approach makes use of the fact that the fermionic path
 integrals in QCD are Gaussian, and can thus be performed analytically.
 The prize paid is the introduction of the fermionic determinant in the
 remaining bosonic path integral.
 The fermionic determinant can be rewritten as a path integral over
 fictitious bosonic ``pseudofermion'' fields.
 A review is given in Ref.~\cite{Wei}.
 The current status of lattice QCD is summarized in
 Ref.~\cite{TouLAT91}.

 \smallskip
 The only exact and practical algorithm (up to now) for numerical
 simulations of field theories involving fermions is the hybrid Monte
 Carlo algorithm \cite{HybridMC}.
 The most time-consuming part in this algorithm is the computation of
 propagators of fermions in background gauge fields.
 Hours of supercomputer time are spent for each matrix inversion
 \cite[p.\,1133]{Sha}.
 It is thus obvious that improved methods for computing fermionic
 propagators are desirable.
 Deterministic MG methods are promising candidates for this task
 because they have proven efficiency in problems of classical
 physics~\cite{BraLAT91}.
 The use of MG methods for quark propagators was proposed by
 Brower, Moriarty, Myers, and Rebbi~\cite{BroMorMyeReb}.

 \smallskip
 Before we turn to embarking on MG methods, we must discuss the
 lattice transcription of fermionic fields.
 The fundamental conceptual problem is that of ``species
 doubling''~\cite{WilsonFermions}.
 A naive discretization of the free fermionic action of one flavor
 leads to $2^d$ degenerate flavors in the classical continuum limit
 (in $d$\ space-time dimensions).
 Actually, the Nielsen-Ninomiya theorem \cite{NieNin} states that on a
 regular lattice there is no local fermion action without doublers which
 possesses chiral symmetry and a positive transfer matrix.
 Smit reviews the problems of lattice fermions in~\cite{SmiReviews}.
 He characterizes them with the words,
 ``Lattice fermions are like a many headed hydra monster''.

 \smallskip
 Two versions of these monsters are in use today:
 Wilson fermions \cite{WilsonFermions} and staggered (or Kogut-Susskind)
 fermions \cite{KogSus,BanSusKog,Sus,ShaThuWei,KluMorNapPet}.
 Wilson's method adds a term to the action which gives masses of the
 order of the cutoff to the unwanted doubler states.
 This procedure, however, breaks the chiral symmetry of the massless
 theory completely.
 Staggered fermions reduce the number of doublers of the naive
 lattice fermion action by a factor of $\zhdh$ while retaining some
 aspects of chiral symmetry.
 Therefore one can use staggered fermions to  study the spontaneous
 breakdown of this remaining lattice symmetry.
 A drawback is, of course,  that one always has $\zhdh$ degenerate
 flavors, whereas there is no restriction on the number of flavors in
 Wilson's formulation (although one has to simulate always an even
 number of flavors in the hybrid Monte Carlo algorithm for technical
 reasons).
 On the other hand, Becher and Joos~\cite{BecJoo} showed that the
 correct formal continuum limit of the staggered Dirac operator is
 equivalent to the Dirac-K\"ahler operator.
 The Dirac-K\"ahler equation is a differential geometric equation for
 $\zhdh$ degenerate flavors already in the continuum, so that there is
 no doubling when one goes to the lattice.

 \smallskip
 In this thesis we will be concerned with staggered fermions,
 mainly for two reasons.
 Firstly, for the merits of staggered fermions just mentioned.
 Secondly, for a computational reason.
 For Wilson fermions successful preconditioning techniques exist
 \cite{Oya,RosDavLep,DeGra,FFT} which are able to accelerate the
 inversion of the fermionic matrix quite substantially.
 However, no such techniques have been proven useful for staggered
 fermions~\cite{RosDav}.

 \smallskip
 We now resume the discussion of computing quark propagators by MG
 methods.
 Traditional methods for inversion of the fermionic matrix suffer
 from CSD when the quark mass is small.
 CSD sets in already for masses larger than physically interesting
 values.
 In connection with deterministic algorithms the term CSD means
 divergence of relaxation times $\tau$ when (inverse) propagators
 approach criticality.
 Roughly speaking, $\tau$ measures the amount of work which is
 necessary to reduce the error of an approximate solution to the
 propagator by a factor of~$e$.
 [A kind of intermediate position between direct solvers and
 (infinite) iteration methods is taken by the conjugate gradient (CG)
 algorithm.
 Although the computational labor increases when the quark mass
 decreases, CG has no well-defined CSD behavior.]

 \smallskip
 The hope is that the problem of CSD can be avoided in an MG approach.
 If one had an efficient MG routine for computing quark propagators
 it could substitute for the conjugate gradient subroutine in
 existing hybrid Monte Carlo programs of full (unquenched) QCD
 and also in quenched approximations.

 \smallskip
 \smallskip
 The application of MG methods to gauge theories is, however,
 not trivial, especially for staggered fermions.
 In particular the meaning of the central notion of smoothness, on
 which MG algorithms are based, needs to be clarified in gauge
 theories.
 It is not a priori clear if MG methods actually work in gauge theories.
 The potential obstacle is the disorder which is inherent in non-Abelian
 gauge fields.
 Edwards, Goodman, and Sokal \cite{EdwGooSok,Edw} pointed out that
 it might be necessary to employ algebraic multigrid (AMG)
 for propagators in gauge fields.
 AMG is not based on a geometric conception of smoothness, but a
 drawback is its more expensive implementation which might make it
 prohibitive.
 Also, the existing theory of AMG does not extend immediately to
 complex matrices.

 \smallskip
 On the other hand, there is an argument that (geometric) MG has a
 chance for propagators in gauge fields.
 The MG procedure presented in this thesis is a gauge covariant
 generalization of a method which is known to be able to eliminate CSD
 in the absence of gauge fields.
 Because of gauge covariance the performance of MG will be the same
 in arbitrary pure gauges.
 However, arbitrary pure gauge field configurations appear to be
 very rough unless they are gauge-fixed.
 Hence, geometric roughness is not necessarily an obstacle for MG\@.

 \smallskip
 Indeed, it will be proved in the present work that {\em the MG method
 works in arbitrarily disordered gauge fields\/}, in principle.
 Unfortunately, the method implemented for this proof is not useful
 for production runs because of computational complexity and storage
 space requirements.
 But it was important to answer questions of principle.

 \smallskip
 \smallskip
 The MG approach to the computation of propagators of staggered
 fermions is not trivial also with respect to another point.
 One of the first questions which has to be answered in MG algorithms
 is the question for the choice of block lattices.
 The answer to this question is obvious in case of bosons, but it is
 not straightforward in case of staggered fermions.
 Different strategies were pursued in the literature
 \cite{BenBraSol,HSVLAT90}.

 \smallskip
 The blocking procedure proposed in the present work starts from the
 requirement that as much as possible is preserved of internal and
 space-time symmetries in the limiting case of vanishing gauge coupling.
 As a result we are forced to choose a scale factor of~$3$
 (or any other odd number).
 Another consequence is the emergence of seemingly overlapping blocks.
 In the limiting case of a pure gauge the blocks have actually no
 sites in common.
 But in nontrivial gauge fields the symmetries of free staggered
 fermions are broken, and for this reason one cannot a priori rule
 out the possibility that the blocks overlap in a nontrivial way.

 \smallskip
 In an MG approach one needs a restriction operator $C$ which averages
 from one grid to the next coarser grid.
 Two qualitatively different proposals will be made for the choice of
 $C$ in case of staggered fermions in nontrivial gauge fields.
 We will call them the ``Laplacian choice'' and the ``Diracian choice''.
 Both proposals reduce to the successful pure gauge construction in the
 limiting case of vanishing gauge coupling.

 \smallskip
 The Diracian choice is preferable for physical reasons, but it is
 computationally quite demanding.
 The reason is that blocks overlap in a nontrivial way, i.\ e.\ every
 fine grid site makes contributions to the block spin (average) at more
 than just one coarse grid site.
 The Laplacian choice is more convenient from the numerical point of
 view.
 It retains the property that also in nontrivial gauge fields the
 seemingly overlapping blocks have actually no sites in common, i.\ e.\
 every fine grid site makes a contribution to the average at exactly
 one coarse grid site.

 \smallskip
 The definition of averaging maps $C$ which will be used in the present
 work has been given the name ``ground-state projection MG''
 \cite{HSVGSP,HSVLAT90,BRVGSP,MacUnpublished}.
 In this method the kernel of (the adjoint of) $C$ is a projector on
 the ground-state of a block-local approximation of a Hamiltonian
 (e.\ g.\ a local approximation of the fermion matrix).
 The idea behind this definition is that the appropriate notion of
 smoothness depends on the dynamics.
 In traditional algorithms the lowest mode of the Hamiltonian is
 responsible for CSD, and this mode should be represented as well as
 possible on coarser grids.
 The ground-state projection choice of $C$ is usable in arbitrary
 space-time dimension $d$ and for arbitrary gauge group.
 An efficient algorithm for the computation of $C$ will be presented.

 \smallskip
 We will prove that the Laplacian proposal for block spins of staggered
 fermions is as ``good'' as the block spin definition for bosons,
 using renormalization group arguments.
 The argumentation uses an ``optimal'' interpolation kernel $\A$
 which is associated with~$C$.
 This $\A$ is a gauge covariant generalization of a kernel which was
 used successfully in rigorous works on constructive quantum field
 theory by \Gaw and Kupiainen \cite{GawKup}.
 The criterion for a good choice of $C$ is exponential decay of $\A$.
 [As a consequence, (exact) effective actions in the sense of the
 renormalization group will remain local.]
 The MG method, which proves that CSD in computations of propagators
 in gauge fields can be eliminated, uses also the generalized
 $\A$-kernel of \Gaw and Kupiainen.

 \smallskip
 Since the Laplacian choice of $C$ defines a good block spin for
 staggered fermions in nontrivial gauge fields, we will restrict
 ourselves to the numerical implementation of this proposal.
 The implementation of the Diracian proposal has to be postponed.

 \clearpage
 \noindent{\sc Organization of the present paper}\newline\indent
 We will investigate the problems which one encounters in making
 MG methods usable for gauge theories in a general framework, and we
 will concentrate on four-dimensional $SU(2)$ gauge
 fields whenever algorithms are tested numerically.
 Generalization to $SU(3)$ would be tedious but not difficult.

 \smallskip
 We begin with a section on deterministic MG methods.
 Here the problem of CSD in computations of propagators is reviewed.
 MG algorithms which were invented to circumvent the phenomenon of CSD
 are introduced.
 The proper way for including gauge fields in MG algorithms is then
 discussed.
 A particularly attractive approach is the ground-state projection
 MG method.
 This method is applicable in arbitrary dimension and for arbitrary
 gauge group.
 It has the advantage that the procedure is gauge covariant, i.~e.\
 no gauge fixing is required.
 The ground-state projection MG method is based on ``block-local
 Hamiltonians''.
 However, the definition of a block-local Hamiltonian requires the
 specification of boundary conditions on block boundaries.
 The necessity of specifying boundary conditions is responsible for the
 fact that ``ground-state projection MG'' is not an a priori defined
 scheme.
 We will choose ``Neumann boundary conditions'', and we will discuss
 this choice in detail.
 The last three subsections of Sec.~\ref{SecMGmethods} will deal with
 the following.
 ``Updating on a multigrid layer consisting of a single site'' will
 be investigated, and gauge covariant related improvements of MG and
 conventional relaxation algorithms are considered.
 None of these improvements needs a tuning of additional parameters.
 We will argue that updating on a single site can be viewed as a global
 rescaling of propagators.
 This procedure respects gauge covariance.
 The section ends with a cautionary remark concerning the term CSD\@.

 \smallskip
 In Sec.~\ref{SecBS} gauge covariant block spins for bosons and
 for staggered fermions are discussed.
 Blocked gauge fields on coarser layers are defined.
 These block spins may not only be used in either deterministic or
 stochastic MG computations, but also in Monte Carlo renormalization
 group studies of gauge theories.
 An efficient algorithm is presented which allows to compute the
 integral kernels needed in ground-state projection MG\@.
 We prove that our proposals for block spins are ``good'',
 using renormalization group arguments.
 In a final paragraph the notion of ``smoothness'' in gauge theories
 is discussed.

 \smallskip
 Then we turn to actual computations of propagators.
 Sec.~\ref{SecFreePropagators} is devoted to the simple case of
 free propagators (this means arbitrary pure gauge) where the bosonic
 and the fermionic problems are equivalent.
 Various MG methods are able to eliminate CSD in this trivial case.

 \smallskip
 Sec.~\ref{SecProof} contains the central result that {\em the MG method
 works in disordered systems\/}.
 MG computations of propagators without CSD in nontrivial gauge
 fields are possible when an ``optimal'' MG scheme is used.
 This statement is true for {\em any} value of the gauge coupling,
 including the case of completely random gauge fields.
 The optimal MG scheme uses gauge covariant generalizations of integral
 kernels whose origin are rigorous works in constructive quantum field
 theory.
 The success of these computations gives ample evidence from the
 deterministic side that Mack's contention concerning the need for
 smooth interpolation kernels is correct.

 \smallskip
 In Secs.~\ref{SecBosonicPropagators} and \ref{SecFermionicPropagators}
 various practical MG algorithms are tested for bosons and staggered
 fermions, respectively, in nontrivial gauge fields.
 The numerical studies reported here are done on lattices of sizes
 up to~$18^4$.
 MG methods give considerable speed-ups compared to conventional
 relaxation algorithms.
 In case of bosons the competitor algorithm (conjugate gradient)
 is outperformed.
 The case of staggered fermions is harder.
 The ``updating on a layer consisting of a single site'' leads to an
 improvement of relaxation algorithms in case of bosons.
 We will discuss with special emphasis on gauge covariance how the
 method can be generalized to the case of staggered fermions.
 However, on lattices up to $18^4$ the method does not pay, and we feel
 unable to predict from numerical experiments whether
 the method pays for staggered fermions on larger lattices of realizable
 sizes, because a volume effect remains with respect to how long it
 takes until errors decay exponentially.

 \smallskip
 \smallskip
 Sec.~\ref{SecSummary} gives a summary and an outlook.
 Two appendices comprise a survey of staggered fermions,
 and a review of the kernels in the optimal MG algorithm together with
 their Fourier representations (in the absence of gauge fields) and
 some tables giving numerical values for integral kernels.

 \smallskip
 Parts of this  thesis have been published before in
 Refs.~\cite{KalNP2,KalMacSpe,KalPL,MacKalPalSpe,KalLastPoint,KalLAT92}.

 \Section{Deterministic Multigrid Methods
 \label{SecMGmethods}}
 \markboth{{\rm\thesection.}\ {\sc Deterministic Multigrid Methods}}{}
 {\sl The content of this section is as follows.
      First the notations for propagators in lattice gauge theories
      are introduced.
      The computation of propagators amounts to solving a very large
      linear system of equations.
      This problem has to be solved frequently in numerical simulations
      of lattice field theories involving fermions (e.\ g.\ QCD).
      In the physically interesting region inverse propagators have an
      eigenvalue which is close to zero.
      The presence of such an eigenvalue makes the use of conventional
      algorithms unpractical, they suffer from critical slowing down.
      Multigrid methods were invented to circumvent this phenomenon.
      Variational multigrid and an ``optimal'' multigrid algorithm are
      discussed.
      The application of multigrid methods to gauge theories requires
      a proper inclusion of the gauge fields.
      A particularly attractive approach is the ground-state projection
      multigrid method.
      This method is applicable in arbitrary dimension and for
      arbitrary gauge group.
      It has the advantage that the procedure is gauge covariant, i.\
      e.\ no gauge fixing is required.
      However, we will see that ``ground-state projection MG'' is not
      defined a priori, and we will discuss what supplements are needed.
      Finally we discuss the special case of ``updating on a multigrid
      layer which consists of a single site'' for bosons.
      (The case of staggered fermions will be discussed in
       Sec.~\ref{SecLPfermi}.)
      Related improvements from a variational point of view will also
      be discussed.
      None of these improvements needs a tuning of additional
      parameters.
      This section ends with a cautionary remark concerning
      the term ``critical slowing down''.}

 \Subsection{Propagators on the Lattice}
 \markboth{{\rm\thesection.}\ {\sc Deterministic Multigrid Methods}}
          {{\rm\thesubsection.}\ {\sl Propagators on the Lattice}}
 {\sl The notations for gauge covariant propagators of bosons and
      of staggered fermions on the lattice are introduced.}
 \smallskip

 \Subsubsection{Notations}
 A propagator in a lattice-regularized field theory is the solution of
 a linear equation
  \begin{equation}
    D \phi = f
  \label{propagator}
  \end{equation}
 on a $d$ dimensional hypercubic lattice $\Lambda$ of sites $z$ and of
 lattice spacing $a$, for given $f$ and non-singular matrix $D$.
 (Later on in the multigrid context, $\Lambda$ will also be called
  $\Lnull$.)
 With respect to numerical simulations of (Euclidean) QCD
 one is interested in fermionic propagators where $D = -\Dirac^2 + m^2$
 \cite{HybridMC,Wei,MacKen}.
 The \rhs $f$ of Eq.~\equ{propagator} is for instance a pseudofermion
 field in simulations with the hybrid Monte Carlo algorithm
 \cite{HybridMC}.
 In case of computations of quark correlation functions, $f$ will be a
 $\delta$-function; then $(-\Dirac + m) \phi$ is the quark propagator.
 As a less complicated problem
 we shall also consider the case of bosons where $D = -\Delta + m^2$.
 The latter is a good toy for tests of new algorithms per se,
 and also because a lot of analytical information is available
 for bosonic propagators.
 $\Delta$ and $\Dirac$ are gauge covariant lattice Laplace or Dirac
 operators, respectively.
 They depend on an external gauge field~$U$.
 In the absence of gauge fields, the Dirac operator is a square root
 of the Laplacian, so that $(-\Dirac^2 + m^2)$ $=$ $(-\Delta + m^2)$
 in this limiting case.

 In the following we will think of $f$ as being a matter field which is
 defined on the sites of $\Lambda$ (e.\ g.\ a pseudofermion field).
 Color indices will always be suppressed, and $\phi (z)$ is an $\Nc
 \times \Nc$ matrix where $\Nc$ is the number of colors.
 Hence, \equ{propagator} is actually an equation for a $(| \Lambda |
 \cdot \Nc \times \Nc )$ matrix~$\phi$.
 Equivalently, one may think of \equ{propagator} as a set of $\Nc$
 systems for $( | \Lambda | \cdot \Nc )$-component vectors.

 \Subsubsection{Discretized covariant Laplace operator}
 We use the standard discretization of the covariant Laplacian where
 $\Delta$ has kernel ($a =$ lattice spacing)
  \begin{equation}
   a^{d+2}\ \Delta ( z_1 , z_2 ) = \left\{ \begin{array}{cl}
 -2d\,\bbbone    & \quad \mbox{if $z_1 = z_2$}\esk \\
  U( z_1 , z_2 ) & \quad \mbox{if $z_1$ n.\ n.\ $z_2$}\esk\\
  0                          & \quad \mbox{otherwise}\esp
                          \end{array} \right.
  \label{LaplaceKernel}
  \end{equation}
 Here $\bbbone$ denotes the $\Nc \times \Nc$ unit matrix,
 ``n.\ n.''\ means nearest neighbor,
 and $U( z_1 , z_2 )$ is the gauge field on the link $(z_1 , z_2)$.
 The oppositely orientated link $(z_2 , z_1)$ carries the gauge field
 $U(z_2 , z_1) = U(z_1 , z_2 )^{\dagger}$, where $^{\dagger}$ denotes
 the Hermitean conjugate of a matrix.

 For a discussion about the lattice transcription of matter and gauge
 fields, and their naive continuum limit we refer to the textbooks of
 Creutz and Rothe \cite{CreBook,Rot}.

 Eq.~\equ{LaplaceKernel} means that $\Delta$ acts on a field $\phi$
 according to
  \begin{eqnarray}
   ( \Delta \phi ) (z) & \equiv & \int_{z'} \Delta ( z , z' )\,\phi (z')
       \equiv a^d \sum_{z'\in\Lambda} \Delta ( z , z' )\,\phi ( z' )
         \nonumber \\  & =  &
   \frac{1}{a^2} \sum_{z'\ \subs{\rm n.n.}\ z}
               \left[ U(z,z')\,\phi (z') - \phi (z) \right] \esp
  \label{Laplace}
  \end{eqnarray}

 \Subsubsection{Discretized covariant Dirac operator}
 In case of fermions we shall use the Euclidean staggered lattice
 formulation \cite{ShaThuWei}.
 In this case the discretized covariant Dirac operator has kernel%
 \footnote{We use the conventions $\eta_{-\mu} = - \eta_{\mu}$
           and $e_{-\mu} = - e_{\mu}$.}
 \begin{equation}
   a^{d+1}\ \Dirac (z_1,z_2) =  \left \{ \begin{array}{ll}
    \eta_{\mu} (z_1)\,U(z_1,z_2)
            & \quad \mbox{ if $z_2=z_1 + \txeh e_{\mu}$,
              $\mu = -d , \ldots , d, \mu\neq 0$},\\
    0       & \quad \mbox{ otherwise} \esk
                        \end{array}  \right.
 \label{DiracKernel}
 \end{equation}
 which means
 \begin{equation}
   (\Dirac \phi ) (z) =
   \frac{1}{a} \sum_{\mu = 1}^d \etamu (z) \left[
               U(z,z+\txeh\emu )\,\phi(z+\txeh\emu )
             - U(z,z-\txeh\emu )\,\phi(z-\txeh\emu )
             \right] \esp
 \label{Dirac}
 \end{equation}
 $\eta_{\mu}$ are the lattice remnants of the $\gamma$-matrices.
 They are complex numbers of modulus 1, and may be chosen as $\eta_1 (z)
 = 1$, $\eta_2(z) = (-1)^{n_1}$, $\eta_3(z) = (-1)^{n_1+n_2}$,
 $\eta_4(z) = (-1)^{n_1+n_2+n_3}$, for $z = \txah (n_1,n_2,n_3,n_4)$.
 Free staggered fermions enjoy discrete translation
 invariance under shifts by twice the separation of neighboring sites.
 Therefore we denote the lattice spacing by $a/2$ in this case.
 $\emu$ denotes a lattice vector of length $a$ in $\mu$-direction.
 The notations and conventions which we use in connection with
 staggered fermions are summarized in Appendix~\theappSF, see also
 Sec.~\ref{SecBSforSF}.

 We note that $\Dirac$ is anti-Hermitean.
 Therefore $(\Dirac + m) \cdot (\Dirac + m)^{\dagger} =
 (-\Dirac^2 + m^2)$.

 \Subsubsection{Gauge covariance}
 An important notion in gauge theories is that of
 {\em gauge covariance}.
 A (local) gauge transformation $g$ is specified on a lattice by a map
 $g: \Lambda \rightarrow G, z \mapsto g(z)$, where $G$ denotes the
 unitary gauge group.
 Under a gauge transformation~$g$ a matter field $\phi$ transforms
 according to
 \begin{eqnarray}
   \phi (z) & \mapsto & \phi' (z) = g(z)\,\phi(z) \esk \nonumber\\
   & &  \label{GTphi} \\
   \phi (z)^{\dagger} & \mapsto & \phi' (z)^{\dagger}
     = \phi(z)^{\dagger}\,g(z)^{-1} \esp \nonumber
 \end{eqnarray}
 The transformation law for a parallel transporter $U(z,w)$ from site
 $w$ to site $z$ is
 \begin{equation}
   U(z,w) \mapsto U' (z,w) = g(z)\,U(z,w)\,g(w)^{-1} \esp
 \label{GTU}
 \end{equation}
 In particular, $U(z,w)$ may be a link variable $U(z_1 , z_2)$.
 (Later on we will encounter general parallel transporters $U(z,w)$
  which are linear combinations of path-ordered products of link
  variables along paths $\C : w \rightarrow z$.)

 {\em The discretized partial differential equation \equ{propagator}
 exhibits gauge covariance\/}, i.\ e.\ if $\phi$ is the solution of
 \equ{propagator} for given $\{ U , f \}$, then $g \phi$ is the solution
 for the gauge-transformed configuration $\{ U' , f' \}$.

 \Subsubsection{Boundary conditions}
 Finally we have to specify boundary conditions for $\Delta$ and
 $\Dirac$, because in numerical computations all lattices have a finite
 extension.
 We shall use periodic boundary conditions in all $d$ directions,
 for gauge fields $U$ and for bosonic and fermionic propagators.
 This kind of boundary conditions is used in present day large scale
 QCD simulations \cite{Bitetal}.

 \Subsection{Critical Slowing Down of Conventional Algorithms
 \label{SecCSD1grid}}
 \markboth{{\rm\thesection.}\ {\sc Deterministic Multigrid Methods}}
          {{\rm\thesubsection.}\ {\sl CSD of Conventional Algorithms}}
 {\sl Classical algorithms for the computation of propagators are
      discussed.
      Nearly critical inverse propagators are of practical interest.
      In nontrivial gauge fields one has to enforce criticality by hand
      (for bosons, and for staggered fermions on relatively small
      lattices), in order to study effects of critical slowing down.
      Criticality is enforced by subtracting the lowest eigenvalue
      $-\mcr$ from the discretized elliptic lattice operator~$D$ and
      adding a small mass term $\Dm$, i.~e.\ we consider bare masses
      $m^2 = \mcr + Dm$.
      In case of bosonic propagators there is a well-known mathematical
      theory about the critical slowing down of traditional relaxation
      methods.
      Due to periodic boundary conditions the critical slowing down
      depends only on the mass but not on the lattice size.
      This point must be especially emphasized:
      {\em One-grid relaxation algorithms for nearly critical bosonic
           propagators are almost not sensitive to the disorder of the
           gauge field, and to the lattice size, as long as equal
           values of $\Dm$ are compared.}
      A kind of intermediate position between direct solvers and
      (infinite) iteration methods is taken by the conjugate gradient
      algorithm which will also be described in this section.
      Although the computational labor increases when the quark mass
      decreases, the conjugate gradient algorithm has no well-defined
      critical slowing down behavior.
      In connection with the conjugate gradient method, we will
      introduce the gauge invariant energy functional which will play
      an important role when one considers algorithms from the
      variational point of view.}
 \smallskip

 \Subsubsection{Problems with the solution of Eq.~\equ{propagator}}
 The lattice operator $D$ in Eq.~\equ{propagator} is a $( |\Lambda |
 \cdot \Nc \times |\Lambda | \cdot \Nc )$ matrix, which is huge in
 practice.
 Therefore direct solvers (e.\ g.\ Gaussian elimination) cannot be
 used.
 One has to employ iterative methods which take advantage of the fact
 that $D$ is very sparse.
 However, the convergence of conventional iterative methods is hampered
 by critical slowing down (CSD).
 In connection with computations of propagators CSD means divergence of
 relaxation times when $D$ approaches criticality,
 i.\ e.\ when it has an eigenvalue which is nearly zero.
 This happens in case of fermions for small quark masses.
 In the bosonic toy model we have to enforce criticality by hand.

 We consider masses $m^2 = \mcr + \Dm$, i.\ e.\ we consider operators
 \begin{equation}
    D = ( - \Delta  + \mcr + \Dm) \quad\mbox{or}\quad
    D = ( - \Dirac^2  + \mcr + \Dm) \esk
 \label{Dcritical}
 \end{equation}
 where $-\mcr$ denotes the lowest eigenvalue of the (semi)positive
 operator $-\Delta$ or $-\Dirac^2$\@.
 $\mcr$ depends on the particular gauge field configuration~$U$.
 However, as all other eigenvalues, it is gauge invariant.
 In nontrivial gauge fields%
 \footnote{A gauge field $U$ is called trivial (or pure gauge) if there
           is a gauge transformation $g$ such that $g(z_1) U(z_1,z_2)
           g(z_2)^{-1} = \bbbone$ for all link variables.
           Therefore this case is equivalent to the absence of gauge
           fields. \label{footnotePG}}
 $-\mcr$ is strictly positive.
 In case of staggered fermions $\mcr$ is a small number which tends to
 zero as the lattice becomes large, but this is not true for bosons.
 Small values of $\Dm$ have to be chosen in order to make $D$ nearly
 critical.
 Critical $D$ are of interest because in practical applications
 $(-\Dirac + m^2)$ will be nearly critical near the continuum limit.

 In this section it will be shown that classical relaxation methods
 for the computation of bosonic propagators exhibit CSD in any
 gauge field configuration.
 This CSD depends only on $m^2$ and {\em not\/} on the lattice size
 $|\Lambda |$.%
 \footnote{For staggered fermions this result will be found
           numerically later on.}
 This is contrarily to the Dirichlet problem, for instance, which is
 discussed in Ref.~\cite{GooSokReview}.
 In our case there is only an implicit dependence on $|\Lambda |$ (and
 $\beta = 2 \Nc\ a^{d-4} / g^2$ where $g$ is the gauge coupling)
 through the value of $\mcr$.
 The dimension $d$ enters in the scaling relation for relaxation times
 only through the constant of proportionality.

 \SubSubsection{Classical Relaxation Algorithms}
 Classical iterative algorithms (see e.\ g.\
 \cite{Var,You,StoBul,GolLoa,HacLGS}) are (damped) Jacobi relaxation
 and SOR (successive over-relaxation).
 SOR includes Gauss-Seidel relaxation as a special case.
 These traditional algorithms generate a sequence $\{ \phi^{(n)}\}$
 of approximate solutions of Eq.~\equ{propagator}.
 Every iteration essentially involves $D$ only in the context of a
 matrix-vector multiplication.

 \smallskip
 Consider for illustration the bosonic problem.
 In this case damped Jacobi relaxation is defined by%
 \footnote{We are a little bit sloppy in writing $(2d + m^2)$ instead of
           $(2d + (ma)^2)$.}
  \begin{equation}
    \phinpo (z) = (1-\omega )\,\phin (z) + \frac{\omega}{2d + m^2}\,%
                [ f(z)\ + \sum_{z'\ \subs{\rm n.n.}\ z} U ( z , z' )\,%
                    \phin ( z' ) ] \esp
  \label{Jacobi}
  \end{equation}
 $\omega$ is called the relaxation parameter.
 It is a real number, equal to 1 for undamped Jacobi iteration.
 SOR makes use of the fact that some updated values for $\phi$ at sites
 $z'$ on the \rhs of \equ{Jacobi} may have been computed before
 when $\phi^{(n+1)} (z)$ is updated.
 Whenever this is the case, $\phin (z')$ in \equ{Jacobi} is replaced by
 $\phinpo (z')$.
 Whether or not $\phinpo (z')$ has already been computed when the
 propagator at site $z$ is updated, depends on the ordering in which
 the sites are swept.
 A scheme which is particularly well-suited for implementations on
 vector computers is the checkerboard ordering.
 Here one employs that $\Delta$ couples only nearest neighbors,
 and that the neighbor of an even site is odd, and vice versa.%
 \footnote{A lattice site $z$ is called even/odd if the sum of its
           integer coordinates is even/odd.}
 Checkerboard SOR reads (modulo exchange even $\leftrightarrow$ odd)
  \begin{eqnarray}
   (i) \mbox{\quad if $z$ is even,}& &   \nonumber \\
     \phinpo (z) & = &  (1-\omega )\phin (z) + \frac{\omega}{2d +m^2}\,%
    [ f(z)\ + \sum_{z'\ \subs{\rm n.n.}\ z} U ( z , z' ) \phin ( z' ) ]
       \esk \nonumber \\   & & \label{checkerboardSOR} \\
   (ii) \mbox{\quad if $z$ is odd,}& & \nonumber\\
     \phinpo (z) & = & (1-\omega )\phin (z) + \frac{\omega}{2d + m^2}\,%
    [ f(z)\ + \sum_{z'\ \subs{\rm n.n.}\ z} U ( z,z' ) \phinpo ( z' ) ]
      \esp \nonumber
  \end{eqnarray}
 SOR with $\omega = 1$ is Gauss-Seidel relaxation.

 \smallskip
 The above iterations can be written as affine fixed point iterations,
 \begin{equation}
  \phi^{(n+1)} = M \phi^{(n)} + (\bbbone - M) D^{-1} f \esp
  \label{iteration}
 \end{equation}
 $M$ is called the iteration matrix.
 Eq.~\equ{iteration} implies that the {\em error\/}
 \begin{equation}
      e^{(n)} \equiv \phi - \phi^{(n)}
  \label{error}
 \end{equation}
 gets (de)amplified by $M$: $e^{(n+1)} = M e^{(n)} = M^{n+1} e^{(0)}$.
 Hence, \equ{iteration} converges if and only if the
 spectral radius $\rho$ of $M$, i.~e.\ the modulus of the largest
 eigenvalue of $M$, is bounded by unity.
 The asymptotic {\em relaxation time\/} $\tau$ is defined by
 \begin{equation}
  \tau = - \frac{1}{\ln \rho (M)}\esp
  \label{tauDef}
 \end{equation}
 Numerically $\tau$ is determined by monitoring ratios $\|r^{(n+1)}\| /
 \|r^{(n)}\|$, which approach $\rho (M)$ asymptotically.
 $r^{(n)}$ denotes the {\em residual\/},
 \begin{equation}
      r^{(n)} \equiv f - D \phi^{(n)}\esp
  \label{residual}
 \end{equation}

 \smallskip
 For any positive definite operator $D$, SOR converges for arbitrary
 initial $\phi^{(0)}$ if and only if $0 < \omega < 2$.
 This is Ostrowski's Theorem \cite[p.\,77]{Var}.
 In our case $D$ is positive definite, so we have an algorithm
 which solves Eq.~\equ{propagator}.
 However, convergence is extremely slow
 when $D$ is nearly critical.
 There is a scaling relation for $\tau$ which reads
 in arbitrary unitary gauge fields
 \begin{equation}
   \tau \propto (\Dm)^{-z/2}  \enspace \mbox{for small} \esk
    \Dm =  m^2 - \mcr \esp
  \label{scalingtau}
 \end{equation}
 irrespective of the lattice size.
 Here $z$ denotes the critical exponent.
 It equals 2 for SOR and damped Jacobi relaxation (with fixed $\omega$).
 Below it will be shown that \equ{scalingtau} is known analytically
 for bosons; its validity for staggered fermions will be found
 numerically later on.

 \smallskip
 \Subsubsection{Bosonic propagators}
 In case of bosonic propagators we can apply a well-known,
 well-elaborated mathematical theory on classical relaxation algorithms.
 This is the theory of so called ``consistently ordered, 2-cyclic''
 matrices \cite{Var,StoBul}.
 A matrix $A$ is called consistently ordered if the eigenvalues of
 the matrix $J( \alpha ) \equiv \alpha\,\mbox{diag}(A)^{-1}L +
 \alpha^{-1}\,\mbox{diag}(A)^{-1}R$ are independent of $\alpha$ for all
 $\alpha \in \bbbc\setminus\{0\}$;
 $\mbox{diag}(A)$ denotes the diagonal of $A$, and $L$ and $R$
 are the lower and upper triangular part, respectively.
 The notion of consistent ordering was introduced by Varga \cite{Var}.
 Criteria for consistent ordering can be found in the literature
 \cite{Var,StoBul,HacLGS}.
 In order to define the term 2-cyclic, we have to introduce the
 (directed) graph $G(A)$, which is associated with $A$.
 If $A$ is an $n \times n$ matrix, $G(A)$ consists of $n$ vertices
 $P_1, \ldots , P_n$, and there is a directed link from $P_i$ to $P_j$
 iff $a_{ij} \neq 0$.
 $A$ is called irreducible if $G(A)$ is connected.
 An irreducible matrix $A$ with nonvanishing diagonal elements
 is called 2-cyclic if the greatest common divisor of the lengths of
 all directed closed paths in $G(J(1))$ equals~2.

 We note that $(-\Delta + m^2)$ is a consistently ordered, 2-cyclic
 operator in arbitrary unitary gauge fields $U$.%
 \footnote{Think of a block structure in terms of the $\Nc\times\Nc$
           blocks $(-\Delta + m^2)(z_1,z_2)$.
           That $(-\Delta + m^2)$ is 2-cyclic can be verified by using
           \cite[Theorem $4.1$]{Var} in connection with
           \cite[Theorems $2.9$, $1.6$]{Var};
           since $(-\Delta + m^2)$ is irreducible, the consistent
           ordering follows from  \cite[Theorem $8.3.11$]{StoBul} in
           connection with  \cite[Theorem $8.3.9$]{StoBul}.}
 Therefore the following conclusions hold \cite{You,Var,StoBul}.
 \begin{itemize}
  \item[$(i)$] The undamped Jacobi iteration converges.
  \item[$(ii)$] Let $\rJ$ be the spectral radius of the Jacobi iteration
    matrix, and $\rGS$ that of the Gauss-Seidel iteration matrix.
    Then $\rJ^2 = \rGS$\,.
  \item[$(iii)$] For SOR there exists an optimal relaxation parameter
     $\oopt$ where the spectral radius $\rom$ of $M$ is minimal:
 \begin{equation}
  \oopt = \frac{2}{1 + \sqrt{1 - \rJ^2}}\quad \esk \quad \quad
  \ropt = \oopt -1 = \left( \frac{\rJ}{1 + \sqrt{1 - \rJ^2}} \right)^2
  \esp
  \label{OptimalOmega}
 \end{equation}
 Generally one has
 \begin{equation}
  \rom = \left\{ \begin{array}{cl}
   \omega - 1 & \mbox{\quad\ for $\oopt \leq \omega \leq 2$\ ,} \\
              & \\
   1 - \omega + \frac{1}{2} \omega^2 \rJ^2 + \omega \rJ
     \sqrt{1 - \omega + \frac{1}{4} \omega^2 \rJ^2}
              & \mbox{\quad\ for $0 \leq \omega \leq \oopt$\,.}
  \end{array} \right.
  \label{rhoSOR}
 \end{equation}
 \end{itemize}

 For $\rJ = 1 - \Dm / (2d + m^2 )$, we obtain
 \begin{equation}
  \oopt = 2 - 2 \sqrt{\frac{\Dm}{d + \frac{1}{2}\mcr}} + O(\Dm )
   \quad \esk \quad \quad
  \tau_{\subs{opt}} = \frac{\frac{1}{2}
                            \left[ d + \frac{1}{2}\mcr \right]^{1/2} }
                           {\left[ \Dm \right]^{1/2} }
  + O\left( [ \Dm ]^0 \right) \esp
  \label{scalingSORopt}
 \end{equation}
 Hence, the critical exponent is 1 for optimal SOR.
 In practice, however, $\oopt$ cannot be determined accurately enough,
 because it depends on the critical mass squared of the particular
 gauge field $U$.
 One would rather choose a fixed $\tilde{\omega} > 1$.
 Eq.~\equ{rhoSOR} implies that $\tau = - 1 / \ln (\tilde{\omega} - 1 )
 = \mbox{constant}$ \label{SORprediction} for all $m^2$ that are larger
 than the value $\tilde{m}^2$ for which $\tilde{\omega}$ is optimal,
 and one has CSD with $z=2$ as in Jacobi relaxation for $m^2 <
 \tilde{m}^2$.
 The scaling relation of the Jacobi algorithm reads
 \begin{equation}
  \tau_J = \frac{\omega^{-1} \left[ 2d + \mcr \right]}{\Dm}
         + ( \omega^{-1} - \txeh ) + O\left( \Dm \right) \esp
  \label{scalingJac}
 \end{equation}
 In anticipation of Secs.~\ref{SecFreePropagators}
 and~\ref{SecBosonicPropagators} we note that the validity of
 \equ{scalingJac}, including the constant $( \omega^{-1} - \txeh )$,
 can be determined numerically to a very high precision.
 In trivial gauge fields Jacobi iteration converges for $0 < \omega
 \leq \omega_{\subs{max}}$, where $\omega_{\subs{max}} = (4d + 2 m^2) /
 (4d + m^2)$ or $\omega_{\subs{max}} = (4d + 2 m^2) / (2d(1+\cos
 \frac{\pi}{N}) + m^2)$ in case that $\Lambda$ is an $N^d$ lattice with
 even or odd $N$, respectively.

 One result of the above considerations must be especially emphasized:
 {\em One-grid relaxation algorithms for nearly critical bosonic
      propagators are almost not sensitive to the disorder of the gauge
      field, and to the lattice size, as long as equal values of $\Dm$
      are compared.}
 At fixed $\beta$ the dependence of $[2d + \mcr ]$ on $\mcr$ can be
 neglected for practical purposes, because $\mcr$ is small compared to
 $2d$, and it does not fluctuate very much from one gauge field
 configuration to the next.
 The dependence of $\mcr$ on the volume of the lattice is also small.
 As a function of $\beta$ or the gauge field correlation length
 the value of $\mcr$ varies only smoothly and slightly.

 \smallskip
 \SubSubsection{Conjugate Gradient Algorithms}
 In the rest of this section we will be concerned with the family of the
 conjugate gradient (CG) algorithms.
 CG algorithms are state of the art in lattice QCD computations
 \cite{ChaKenRow,RosDavLep,MacKen,HenSetDav,BurIrv,KenJulich,%
       TFproject,Bitetal,Columbia}.
 Hence, any new algorithm must be compared with CG.
 (The UKQCD group \cite{UKQCD} uses an over-relaxed minimal residual
  algorithm, though.)

 General CG methods are designed for finding stationary points
 of continuously differentiable functions \cite{Pol,Preetal}.
 The CG algorithm for propagators starts from the fact that solving
 Eq.~\equ{propagator} is equivalent to minimizing the energy functional
  \begin{equation}
     K [ \phi ]\ =\ \frac{1}{2} < \phi , D \phi >  - < \phi , f >
  \label{EnergyFunctional}
  \end{equation}
 where the scalar product is given by
  \begin{equation}
    < \phi , f >\ =\ \frac{1}{|\Lambda |} \sum_{z\in\Lambda}
      \frac{1}{\Nc}\,\re\Tr\left[ \phi (z)^{\dagger} f (z) \right] \esp
  \label{ScalarProduct}
  \end{equation}
 \mbox{}

 \begin{quotation}
 {\small\bf\noindent Gauge invariance property of the energy
                     functional.}
 \newline\indent
 {\em We note that the energy functional $K$ is gauge invariant,
      i.~e.\ $K[ \phi ] = K[ g\,\phi ]$.}
 (Remember that we assumed that $f$ transforms like a matter field
  under gauge transformations.)
 Later on we will encounter situations where $f(z) \mapsto g(z)\,f(z)\,%
 g(w)^{-1}$ and $\phin (z) \mapsto g(z)\,\phin(z)\,g(w)^{-1}$ under
 gauge transformations, where $w$ is fixed.
 {\em Note that $K$ is also gauge invariant in these situations.}
 \end{quotation}

 The CG method is iterative in the sense that starting with an
 arbitrary initial $\phi^{(0)}$, it yields a sequence
 $\phi^{(0)} \rightarrow \phi^{(1)} \rightarrow \phi^{(2)} \cdots$
 which converge to the solution of Eq.~\equ{propagator}.
 But unlike the relaxation methods discussed above, CG arrives at the
 solution after at most $| \Lambda |$ steps (provided the
 arithmetic is exact).

 The CG algorithm is a steepest descent method which performs a
 $(k\!+\!1)$ dimensional minimization in the step $\phi^{(k)}
 \rightarrow \phi^{(k+1)}$.
 $\phi^{(k+1)}$ is determined such that
 \begin{equation}
  K[ \phi^{(k+1)} ] = \min_{\alpha_0 , \ldots , \alpha_k \in \bbbc}
  K[ \phi^{(k)} + \alpha_0 r^{(0)} + \cdots + \alpha_k r^{(k)} ] \esp
  \label{MinimizationCG}
  \end{equation}
 The residuals $r^{(i)}$, $i = 0, \ldots , k$, are orthogonal, and thus
 independent, as long as $r^{(k)} \neq 0$.
 Since at most $| \Lambda |$ vectors%
 \footnote{Remember that, strictly speaking, the vectors here are
           actually matrices.}
 are independent, there must be an $l \leq | \Lambda |$ with $r^{(l)} =
 0$ and $\phi^{(l)}$ solves Eq.~\equ{propagator}.

 \smallskip
 \noindent
 The cookbook recipe of the CG method is the following
 \cite{StoBul,HacLGS,GolLoa}.
 \begin{itemize}
  \item[(1)] Choose any $\phi^{(0)}$ and set $p^{(0)} = r^{(0)}
             = f - D \phi^{(0)}$, \quad $k=0$.
  \item[(2)] If $r^{(k)} = 0\ \Leftrightarrow\ p^{(k)} = 0$ : STOP,
             $\phi^{(k)}$ is the solution of $D\phi = f$.
  \item[(3)] Else compute
   \begin{equation}
   \begin{array}{rclrcl}\protect\d
    a_k = \frac{<r^{(k)}\,,\,r^{(k)} >}{<p^{(k)}\,,\,D p^{(k)} >}
    & \equiv &\d \frac{<r^{(k)}\,,\,p^{(k)} >}{<p^{(k)}\,,\,D p^{(k)} >}
         \esk &
    \quad \d \phi^{(k+1)} & = & \d \phi^{(k)} + a_k\,p^{(k)}\esk \\
     & & & & & \\
    \d r^{(k+1)} & = & \d r^{(k)} - a_k\,D p^{(k)}  \esk     & \quad
    \d b_k & = &\d \frac{<r^{(k+1)}\,,\,r^{(k+1)} >}
    {<r^{(k)}\,,\,r^{(k)} >} \esk \\
     & & & & & \\
    \d p^{(k+1)} & = & \d r^{(k+1)} + b_k\,p^{(k)} & &
   \end{array}
   \label{CG}
   \end{equation}
  \item[(4)] Increase $k$ by 1 and go to (2).
 \end{itemize}
 See Refs.\ \cite{StoBul,HacLGS,GolLoa,WilRei} for properties of the
 $\phi^{(k)}$, $p^{(k)}$, $r^{(k)}$, for a proof of convergence of the
 CG method and for a discussion about its numerical properties.

 \smallskip
 In practice one will inevitably have to deal with round-off errors,
 and one will stop the algorithm when $\| r^{(k)} \|$ is ``small
 enough'' (e.\ g.\ when $r^{(0)}$ is reduced by $10^{-5}$).
 The same stopping criterion will of course be used in relaxation
 algorithms which never reach the solution otherwise.
 The reason for the success of CG is that the stopping criterion is
 reached in much less than $| \Lambda |$ iterations.

 \smallskip
 Sometimes one hears the statement that CG has a critical exponent
 $z \approx 1$ or even the statement $z=1$ for CG\@.
 This must be taken cautiously.
 Strictly speaking, CG has no well defined $z$ because it is not
 an infinite process.
 If one nevertheless wants to define an asymptotic relaxation time,
 $\tau$ would be zero, because after a finite number of iterations the
 error is zero.
 Regarded from that point of view, CG is a direct solver.
 What causes people to say ``$z \approx 1$ for CG'' is the following.
 One can prove the bound~\cite{HacLGS}
 \begin{equation}
   \| D^{1/2}\,e^{(k)} \|\ \leq\ \frac{2\,c^k}{1 + c^{2k}}\,%
   \| D^{1/2}\,e^{(0)} \| \quad \mbox{with} \quad
    c = \frac{\sqrt{\kappa}-1}{\sqrt{\kappa}+1} \esp
  \label{ConvergenceCG}
  \end{equation}
 Here $\kappa$ denotes the condition number of $D$, i.\ e.\ the
 ratio of the largest to the smallest eigenvalue.
 If CG were an infinite iteration and if the bound were stringent, one
 would derive the scaling relation \equ{scalingtau} from
 \equ{ConvergenceCG} with $z=1$.
 But even if the iteration were infinite, the situation is more
 complicated, see the discussion in Ref.~\cite{HacLGS}.

 \smallskip
 Finally we note that preconditioning techniques \cite{GolLoa,HacLGS}
 exist which are very successful in accelerating the computation of
 propagators of Wilson fermions \cite{Oya,RosDavLep,DeGra,FFT}.
 However, no such techniques have been proven useful for staggered
 fermions~\cite{RosDav}.
 For this reason we will not discuss and use preconditioning in the
 present thesis.

 \Subsectiontocentry{Elimination of Critical Slowing Down by Multigrid
                     Algorithms}
                    {Elimination of Critical Slowing Down \protect\\
                     by Multigrid Algorithms
 \label{SecMGalgorithms}}
 \markboth{{\rm\thesection.}\ {\sc Deterministic Multigrid Methods}}
          {{\rm\thesubsection.}\ {\sl Elimination of CSD by MG
                                      Algorithms}}
 {\sl The general deterministic multigrid method is explained in this
      subsection.}
 \smallskip

 Multigrid (MG) algorithms were invented to circumvent the problem of
 CSD in the solution of discretized partial differential equations
 (in the absence of gauge fields).
 The slowness of traditional algorithms is overcome by updates on
 various length scales.
 Introductions to this subject can be found in the classical papers
 of Brandt, and St\"uben and Trottenberg \cite{MGclassical}, in the
 textbook of Hackbusch \cite{HacMG}, at a less advanced level in the
 books of McCormick \cite{McCor} and of Hackbusch \cite{HacLGS},
 and at a very elementary level in the book of Briggs \cite{Bri}.

 The basic observation for MG methods is the following.
 Classical relaxation algorithms are effective in smoothing the
 error, but as soon as the error is smooth (on length scale $a$)
 it is reduced only very slowly because of CSD.
 However, a smooth function can be represented very well on a coarser
 lattice.
 Suppose for instance that the values of a lattice function
 are given only on every second site.
 Then, if one knows that the function is smooth, one can reconstruct
 it to a good accuracy by interpolation.
 We will now explain these ideas in more detail.
 (The reader may prefer not to think of the presence of gauge fields
  in Eq.~\equ{propagator} until Sec.~\ref{SecInclusionGF}.)

 In the MG approach for solving Eq.~\equ{propagator}
 we divide the original hypercubic lattice $\Lambda$ of lattice spacing
 $a$ into hypercubes (``blocks'') $x$ consisting of $L_b^d$ sites
 $z \in \Lambda$, with $L_b \in \bbbn$,%
 \footnote{The requirement $L_b \in \bbbn$ is not compulsory; see
           e.\ g.\ Sec.~\ref{SecSqrt3}.}
 typically $L_b = 2 , 3$.
 We identify each such hypercube $x$ with the site $\hat{x}$ at its
 center, and we write $z \in x$ if $z$ is a site in block $x$.
 (If $L_b$ is even there is no such distinguished center. Then define
  arbitrarily an $\hat{x}$ in one block; this defines the other block
  centers by the requirement that the block lattice is regular.)
 The sites $\hat{x}$ form the first block lattice $\Lone$ with
 lattice spacing $L_b a$, and so on.
 This yields a sequence of lattices $\Lambda = \Lnull$, $\Lone$,
 $\Ltwo$,~$\ldots$ of increasing lattice spacing $a_i$, viz.\ $a_{i+1}
 = \L a_i$ with $a_0 = a$.
 (One may also use different blocking factors $\L$ on different layers
  of the MG).

 After some relaxation sweeps on $\Lnull$ one gets an approximation
 $\phin$ to $\phi$ which differs from the exact solution by an
 error $e^{(n)} \equiv e_0$, Eq.~\equ{error}.
 The error satisfies the {\em residual equation}
  \begin{equation}
    D_0 e_0 = r_0 \esk
  \label{residualEq}
  \end{equation}
 where $r_0 \equiv r^{(n)}$ is the residual, \equ{residual}, and we
 wrote $D_0$ for $D$.
 If $e_0$ is smooth, it is determined to a very good accuracy by a
 function $e_1$ on the next coarser lattice $\Lone$, and can be
 represented in the form
 \begin{equation}
  e_0 = \A e_1
 \label{interpolatedError}
 \end{equation}
 with an interpolation map $\A$ which should be so chosen that it maps
 functions on $\Lone$ into smooth functions on $\Lnull$.
 Conversely, $e_1$ can be obtained from $e_0$ with the help of an
 averaging map $C$ which satisfies
 \begin{equation}
  C \A = \bbbone \esp
 \label{CAequalsOne}
 \end{equation}
 It follows that $e_1 = C e_0$.
 Inserting Eq.~\equ{interpolatedError} into Eq.~\equ{residualEq} and
 acting on the result with $C$, we see that $e_1$ will satisfy the
 equation
 \renewcommand{\theequation}{\thesection.\arabic{equation}\alph{abc}}
 \setcounter{abc}{1}
 \begin{equation}
  D_1 e_1 = r_1
 \label{residualEQcoarse}
 \end{equation}
 with
 \addtocounter{abc}{1}\addtocounter{equation}{-1}
 \begin{equation}
  D_1 = C D_0 \A \esk \quad r_1 = C r_0 \esp
 \label{CoarseGridOperator}
 \end{equation}
 \renewcommand{\theequation}{\thesection.\arabic{equation}}
 The problem has been reduced to an equation on the coarser lattice
 $\Lone$ which has fewer points.
 If there is still too much CSD at this level,
 one may repeat the procedure, going to coarser and coarser lattices.
 The procedure stops, because an equation on a ``lattice'' $\Lambda^N$
 with only a single point is easy to solve.

 After solution of Eq.~\equ{residualEQcoarse} one replaces $\phin
 \mapsto \phint \equiv \phin + \A e_1$.
 Note that the residual of the corrected approximation
 $\phin + \A e_1$ vanishes when it is transfered back to $\Lone$.
 If $\A e_1$ were equal to $e_0$, then $\phint$ would be the solution
 of Eq.~\equ{propagator}.
 In practice, however, one has to repeat the procedure:
 do relaxation with $\phint$, solve the residual equation for the new
 error, etc.

 The reason for the efficiency of the MG method is that with a
 suitable choice of $C$, $\A$, $D_1$ etc.\ only a few iterations are
 needed to reduce the error to a small value, independent of the
 mass term $m^2$.
 In other words, CSD is completely eliminated by MG\@.
 This statement has been known to be true in the absence of gauge
 fields, and it is the subject of this thesis to address the question
 what the effects of the inclusion of gauge fields are.
 A crucial problem is how to define and exhibit smooth functions in
 a disordered context, i.\ e.\ when translation symmetry is strongly
 violated.

 Another advantage of the MG method is that the computational labor
 for one MG iteration is comparable to that of conventional relaxation,
 irrespective of the total number of layers.
 For details of this work estimate see
 \cite{MGclassical,HacMG,Bri,GooSokReview,HacLGS}.

 There are further terms which are relevant in MG algorithms
 for which we refer to the literature.
 These terms include the ``cycle control parameter $\gamma$'', the
 notion of ``V-cycles'' ($\gamma = 1$) and ``W-cycles'' ($\gamma = 2$),
 etc.
 We will not need these terms.
 We are interested in four space-time dimensions where the volumes of
 subsequent lattices differ by a factor of $L_b^4$.
 Throughout this thesis we will always fix $L_b = 3$ without further
 notice.
 Thus, $\Lone$ has 81 times fewer points than $\Lnull$.
 Eq.~\equ{residualEQcoarse} on the coarser lattice was solved exactly
 by CG.%
 \footnote{This two-grid algorithm is equivalent to an MG algorithm
           with $\gamma = \infty$ (or a large number).}
 This suffices to test the power of the MG method.

 Finally we note that the MG iteration is also an affine fixed point
 iteration like \equ{iteration}.
 For instance, the iteration matrix $\M$ of a two-grid algorithm
 is given by $\M = M^{\nu_2} \left[ \bbbone - \A ( D_1 )^{-1} C D_0
 \right] M^{\nu_1}$, where $\nu_1$ and $\nu_2$ denote the number of
 relaxation sweeps on $\Lnull$ before and after the MG correction step,
 respectively.
 In actual computations only the sum $\nu_1 + \nu_2$ matters, and we
 chose $\nu_1 + \nu_2 = 1$.
 By using the Fourier representations of integral kernels it is an easy
 exercise to show that $\M$ annihilates the lowest mode of $D_0$
 with different choices of $C$ and $\A$; see Appendix~\theappKernels.2.

 \Subsection{Variational Multigrid
 \label{SecVariationalMG}}
 \markboth{{\rm\thesection.}\ {\sc Deterministic Multigrid Methods}}
          {{\rm\thesubsection.}\ {\sl Variational MG}}
 {\sl The variational choice of averaging and interpolation operators
      in MG algorithms is introduced.}
 \smallskip

 In order to specify an MG algorithm, we have to make a specific choice
 for the restriction operator $C$ and for the interpolation operator
 $\A$.
 These operators will be defined by their integral kernels $C (x,z)$
 and $\A (z,x)$.
 ($z \in \Lambda^j$, $x \in \Lambda^{j+1}$, e.\ g.\
  $z \in \Lnull$, $x \in \Lone$)
 The notations in this thesis will always be such that $z$ denotes
 sites in the finer lattice and $x$ denotes sites in coarser layers.

 For reasons of practicality one must require that
 \begin{equation}
  \A (z,x) = 0 \quad \mbox{unless $z$ is near $\hat{x}$.}
 \label{practicality}
 \end{equation}
 Two special choices for $C$ and $\A$ will be discussed in this and
 the next section.

 \smallskip
 {}From the variational point of view one requires that the energy
 functional $K$ \equ{EnergyFunctional} should be lowered as far as
 possible in every MG correction step $\phin \mapsto \phin + \A e_1$.
 It follows that the averaging map $C$ and the interpolation map $\A$
 are adjoints of one another \cite{MGclassical,HacMG,GooSokReview}:
  \begin{equation}
    C = \A^{\ast} \esk
  \label{CequalsAstar}
  \end{equation}
 and that the coarse grid operator $D_1$ is defined as
  \begin{equation}
    D_1 = C D_0 \Cstar \equiv C D \Cstar  \esp
  \label{Galerkin}
  \end{equation}
 This is called the Galerkin definition of $D_1$.
 The integral kernel of the adjoint averaging operator is
 $\Cstar (z,x) = C(x,z)^{\dagger}$.

 Eq.~\equ{CAequalsOne} imposes the normalization condition $C \Cstar
 = \bbbone$ on the averaging operator~$C$ in variational MG\@.

 \Subsection{``Optimal'' Multigrid
 \label{SecOptMG}}
 \markboth{{\rm\thesection.}\ {\sc Deterministic Multigrid Methods}}
          {{\rm\thesubsection.}\ {\sl ``Optimal'' MG}}
 {\sl An ``optimal'' MG scheme is introduced.
      In this scheme there is complete decoupling between the layers
      of the MG\@.
      This decoupling will also hold in the presence of gauge fields.}
 \smallskip

 Given the averaging kernel $C$, there exists an ideal choice
 of the interpolation kernel $\A$.
 It is determined as follows.
 For every function (``block spin'') $\Phi$ on $\Lone$, $\phi = \A \Phi$
 minimizes the action $\H\ =\ < \phi\,,\,D\,\phi >$
 subject to the constraint $C \phi = \Phi$.
 With this choice of $\A$, $D_1$ is guaranteed to be self-adjoint.
 A good ``choice of block spin'', i.\ e.\ of $C$, is characterized
 by the fact that the ideal kernel $\A (z,x)$ associated with it has
 good locality properties.
 This means that $\A (z,x)$ is big for $z \in x$, and decays
 exponentially in $| z - \hat{x} |$ with decay length one block lattice
 spacing.

 The above characterization of $\A$ is equivalent to saying that
 with the ideal choice of $\A$, there is complete decoupling between
 layers.
 This means that the action $\H$ can be written as a sum of actions
 for the different layers of the MG, viz.\
 $\H = \sum_j \H_j$, with $\H_j\ =\newline
                                    < \zeta_j\,, D_j\,\zeta_j >$\@.
 $\zeta_j$ is a field which is defined on $\Lambda^j$, and $D_j$ is
 the effective inverse propagator on $\Lambda^j$.
 In other words:
 Complete decoupling between layers means that the propagator is a sum
 of contributions from layers, and these contributions from different
 layers satisfy effective difference equations which do not couple.
 As a result, the convergence speed is determined by the convergence
 speed on the individual layers.
 We note that complete decoupling $\H = \sum_j \H_j$ does not hold
 in variational MG\@.

 For the purpose of numerical computations, it is convenient to
 determine the optimal $\A$ as solution of the equation
 \begin{equation}
  \left( [ D + \kappa\,\Cstar C ] \A \right) (z,x)
  = \kappa\,\Cstar (z,x)
 \label{optimalA}
 \end{equation}
 for large $\kappa$.
 The solution of Eq.~\equ{optimalA} yields the same $\A$-kernel which
 was proposed in Mack's Carg\`{e}se lectures \cite{MacCargese}
 for use in ``optimal'' MG Monte Carlo simulations.
 In Ref.~\cite{MacCargese}, $\A$ was determined for 1-component $\phi^4$
 theory by a relaxation method.
 This is feasible, because in the absence of gauge fields and for the
 step function kernel $C$ of Eq.~\equ{CwithoutGF} below, $(\Delta \A
 )(z,x)$ are constants on blocks $x$ as functions of $z$.
 Another possible computation of $\A$ would be by standard optimization
 algorithms, making use of the above characterization of $\A$ as
 solution of an extremization problem.

 In the presence of gauge fields, $\A$ is computed as the solution
 of the gauge covariant generalization of Eq.~\equ{optimalA}.
 We note that complete decoupling between the layers of the MG will
 continue to hold in the presence of gauge fields.

 \smallskip
 The origin of the optimal $\A$ lies in the works \cite{GawKup}
 of \Gaw and Kupiainen in constructive quantum field theory,
 see also their Les Houches lectures \cite{GawKupLH}.
 \Gaw and Kupiainen gave rigorous proofs of the existence of the
 continuum limit of some lattice field theories without gauge fields,
 using block spin renormalization group methods.
 The use of the Gaw\c{e}dzki-Kupiainen kernels as a starting point
 for numerical work was proposed by Mack \cite{MacCargese}.
 He pointed out that it will be essential for beating CSD in interacting
 models that the layers of an MG decouple as much as possible.
 A necessary condition for this is smoothness of $\A$.
 The correctness of Mack's contention was confirmed in an impressive
 way by Hasenbusch and Meyer \cite{HasMeyMac,HasMey}.

 A derivation of the optimal $\A$, including Eq.~\equ{optimalA},
 is summarized in Appendix~\theappKernels\@.
 In Secs.~\ref{SecOptA} and~\ref{SecProof}
 further properties of the ideal $\A$ will be discussed.
 By its use it was possible to demonstrate that the MG method can cope
 with the frustration which is inherent in non-Abelian gauge fields
 (Sec.~\ref{SecProof}).
 Unfortunately, the optimal $\A$ does not fulfill the practicality
 condition \equ{practicality}, so that the idealized MG algorithm
 cannot be used for production runs.
 But the results of Sec.~\ref{SecProof} settle questions of principal
 importance.

 \Subsection{Inclusion of Gauge Fields
 \label{SecInclusionGF}}
 \markboth{{\rm\thesection.}\ {\sc Deterministic Multigrid Methods}}
          {{\rm\thesubsection.}\ {\sl Inclusion of Gauge Fields}}
 {\sl The requirements for making MG algorithms covariant in the
      presence of gauge fields are discussed.
      When an algorithm is gauge covariant, the necessity for fixing a
      gauge in computation of propagators is avoided.}
 \smallskip

 Now we turn to the point of including gauge fields properly
 into the MG algorithm.
 This should be done in such a way that gauge covariance is ensured.
 Then no gauge fixing is required in computations of propagators.
 Note that Jacobi relaxation \equ{Jacobi}, SOR \equ{checkerboardSOR}
 and CG \equ{CG} are gauge covariant in the sense that all $\phin$ are
 gauge transformed by $g$ if $g$ is applied before relaxation is
 started.
 (We always begin with an initial $\phi^{(0)} = 0$.)
 The gauge covariance property is to be retained in multigrid
 algorithms.

 The block spin $\Phi = C \phi$ should transform under gauge
 transformations \equ{GTphi}, \equ{GTU} like a matter field sitting at
 block centers $\hat{x}$.
 Similarly, the ``background field'' $\psi = \A \Phi$ should transform
 like $\phi$.
 This is achieved if under gauge transformations
 \begin{eqnarray}
      \A (z,x) &\mapsto& g(z)\,\A (z,x)\,g(\hat{x})^{-1}\esk \nonumber\\
       & & \label{CovKernels} \\
        C (x,z) &\mapsto& g(\hat{x})\,C (x,z)\,g(z)^{-1} \esp \nonumber
 \end{eqnarray}
 Then the equations of Secs.~\ref{SecMGalgorithms}\ --\ \ref{SecOptMG}
 become gauge covariant.
 Eq.~\equ{CovKernels} is consistent with $C\A = \bbbone$.
 Note that the transformation law \equ{CovKernels} is also consistent
 with the variational choice $\A = \Cstar$.

 The most general expression of a kernel $\A$ with covariance
 property \equ{CovKernels} is a weighted sum of parallel transporters
 $U( \C )$ along paths $\C$ from $\hat{x}$ to $z$, i.\ e.\
 \begin{equation}
    \A (z,x) = \sum_{\C\ :\ \hat{x} \mapsto z} \varrho( \C )\ U( \C )
    \esk
 \label{SumofPaths}
 \end{equation}
 where $\varrho( \C )$ are numbers.
 And analogously for $C$.
 One sees from \equ{SumofPaths} that the integral kernels of $C$ and
 $\A$ are $\Nc \times \Nc$ matrices in gauge theories.
 However, they will not be elements of the gauge group, in general.
 {}From the fact that $C$ and $\A$ are matrices, it follows especially
 that there is a color-mixing in averaging and in interpolation.

 An MG approach which is based on the representation~\equ{SumofPaths}
 is the ``parallel transported multigrid'' (PTMG) method, introduced by
 Ben-Av, Brandt, and Solomon \cite{BenBraSol} for computations in the
 Schwinger model (two-dimensional QED).
 The PTMG approach is also described in Ref.~\cite{HaretalLAT90},
 in Ben-Av's thesis \cite{Benthesis}, and in \cite{PTMG_GMD}.

 We will prefer not to specify the weights $\varrho$ in \equ{SumofPaths}
 explicitly, but to determine $\A$ and $C$ as solutions of covariant
 equations.
 This alternative approach is the ``ground-state projection MG'',
 which will be discussed next.

 \Subsection{Ground-State Projection Multigrid
 \label{SecGSP}}
 \markboth{{\rm\thesection.}\ {\sc Deterministic Multigrid Methods}}
          {{\rm\thesubsection.}\ {\sl Ground-State Projection MG}}
 {\sl The particularly attractive ``ground-state projection multigrid''
      approach is defined.
      This method is applicable in arbitrary dimension and for arbitrary
      gauge group.
      It has the advantage that the procedure is gauge covariant, i.~e.\
      no gauge fixing is required.
      The ground-state projection MG method is based on ``block-local
      Hamiltonians''.
      However, the definition of a block-local Hamiltonian requires the
      specification of boundary conditions on block boundaries.
      The necessity of specifying boundary conditions is responsible
      for the fact that ``ground-state projection MG'' is not an a
      priori defined scheme.
      We will choose ``Neumann boundary conditions'', and we will
      discuss this choice in detail.}
 \smallskip

 The central idea of the ground-state projection MG philosophy is that
 a local action should define the block spin (or $C$, respectively).
 The averaging operator $C$ from a grid to the next coarser grid is a
 projector on the ground-state of a block-local Hamiltonian.
 The adjoint of $C$ satisfies a gauge covariant eigenvalue equation,
 Eq.~\equ{EVequationC} below.
 The solution of the eigenvalue equation is made unique by imposing
 a normalization and a covariance condition (Eqs.~\equ{normalizationC}
 and \equ{CovarianceCondition}).
 The idea behind a ground-state projection definition is that the
 appropriate notion of smoothness depends on the dynamics, i.~e.\ on
 $D$, in general.
 Results, which will be reported in later sections, confirm the
 insight that {\em smooth means little contributions from
 eigenfunctions to high eigenvalues of $D$.}
 This point is important in systems in gauge fields and for other
 disordered systems.

 \SubSubsection{Averaging Map $C$ for Free Bosons}
 Let us first discuss the averaging map $C$ in the absence of gauge
 fields, and for bosons; the case of staggered fermions will be
 treated in Sec.~\ref{SecBSforSF}.
 It is reasonable to demand that $C$ is local in the sense that
 the averaged field (``block spin'') $\Phi (x) = (C \phi )(x)$ at a
 block lattice site $x$ receives contributions only from $\phi (z)$ for
 $z \in x$, i.\ e.\ $C(x,z) = 0$ unless $z \in x$.
 The natural choice inside (non-overlapping) blocks is a constant.
 Hence,%
 \footnote{A remark about scale factors is in order here.
           The operator $C$ is dimensionless, but its integral kernel
           has dimension of a length$^{-d}$.
           Almost always factors of $(L_b a)^{-d}$ are neglected in
           other works.
           But we stress that one has to keep track of them, otherwise
           there will appear inconsistencies.
           One can set the lattice spacing $= 1$ on each layer
           separately.
           But then one has to watch out for factors $\L$ when one
           goes from one layer to the next.
           \setcounter{fnmarkone}{\value{footnote}}}
  \begin{equation}
   C (x,z) =  \left\{ \begin{array}{cl}
  (L_b a)^{-d} \cdot 1 & \quad \mbox{if $z \in x$}\esk \\
     0                 & \quad \mbox{otherwise}\esp \end{array} \right.
  \label{CwithoutGF}
  \end{equation}
 With this choice, $\Phi$ is the block average of $\phi$,
  \begin{equation}
   \Phi (x) = \int_{z} C(x,z) \phi (z)
     = L_b^{-d} \sum_{z \in x} \phi (z) \equiv \av{z \in x} \phi (z)\esp
  \label{BlockAverage}
  \end{equation}
 This block spin transformation was also used by \Gaw and Kupiainen
 \cite{GawKup}.
 We will now see how it can be naturally generalized so that gauge
 fields are incorporated.

 \SubSubsection{$C$ as the Solution of an Eigenvalue Equation}
  We recall that we denoted by $\Cstar (z,x)$ the integral kernel
 of the adjoint of $C$\@.
 One has the equality $\Cstar (z,x) = C(x,z)^{\dagger}$, where
 $^{\dagger}$ denotes Hermitean conjugation of a matrix, as usual.
 The adjoint of the averaging kernel \equ{CwithoutGF} is a solution of
 the eigenvalue equation
  \begin{equation}
  ( -\LAPN \Cstar ) ( z , x )  = \lambdanull\,\Cstar (z,x) \esk
  \label{EVequationC}
  \end{equation}
 together with the subsidiary condition $\Cstar (z,x) = 0$ if $z \not\in
 x$.
 $-\LAPN$ is the negative lattice Laplacian with Neumann boundary
 conditions on the boundary of block $x$ (see below), and $\lambdanull$
 is its lowest eigenvalue.
 $\LAPN$ acts on argument $z$.
 In the absence of gauge fields $\lambdanull$ equals zero for all
 blocks~$x$.
 In this case solutions of \equ{EVequationC} are constants on blocks.
 These constants can be determined by the normalization condition $C
 \Cstar = \bbbone$, which reads for the kernel%
  \begin{equation}
  (C \Cstar ) ( x_1  , x_2 ) = \int_z C(x_1,z) \Cstar (z,x_2)
  = \delta ( x_1 - x_2 ) \equiv (L_b a)^{-d} \delta_{x_1,x_2} \esp
  \label{normalizationC}
  \end{equation}

 Following Mack \cite{MacUnpublished}, this procedure can be
 re-interpreted as follows.
 Define a notion of block-local frequency which depends only on
 the behavior of the function which is to be decomposed into
 frequency components on the chosen block $x$.
 Thus, define frequency squared as eigenvalue
 of the negative Laplacian with Neumann boundary conditions.
 Define the block spin in two steps.
 First define the lowest frequency part $\psi (z)$ of $\phi (z)$
 by projection
 \begin{equation}
  \psi (z) = \lim_{t \rightarrow \infty}
  ( \exp \left[ - t ( -\LAPN - \lambdanull ) \right] \phi )(z) \esp
  \label{projection}
 \end{equation}
 $\psi (z)$ will be a smooth function of $z$ inside the block.
 In the second step the block spin is defined equal to the value
 of this smooth function at the block center $z = \hat{x}$.

 We note that Neumann boundary conditions on the Laplacian yield a local
 approximation of $\Delta$ which preserves the invariance of the action
 $\H_0\ =\ < \phi , -\Delta \phi >$ under shifts by constant fields.
 One sees that the block spin $\Phi (x)$ retains the component of
 $\phi$ associated with the lowest eigenvalue of $-\LAPN$.

 For block spins in (non-Abelian) gauge theories one can proceed in
 exactly the same way, defining $\Cstar$ as solution of
 Eq.~\equ{EVequationC}.
 Now $C$ depends on $U$, though we will not indicate that explicitly.
 $\LAPN$ will be a discrete substitute for the Laplace operator
 with Neumann boundary conditions on the continuum.

 The idea is that the definition of block spins involves
 dynamical information --- think of $\int_{z\in x} \txeh \Nc^{-1}\,%
 \re\Tr[\phi (z)^{\dagger}(-\LAPN \phi)(z)]$ as the part of the kinetic
 energy which is associated with the inside of block $x$.
 Thus, what is called ``low frequency'' is actually determined by
 kinetic energy.
 One could also admit dielectric gauge fields $U$ \cite{MacDLGT},
 especially on coarser layers.

 \SubSubsection{Neumann Boundary Conditions on the Lattice}
 Neumann boundary conditions (b.~c.)\ specify the values of the normal
 gradients of a function $\phi$ on a boundary.
 In numerical analysis the term ``Neumann boundary conditions'' is used
 synonymously with ``homogeneous Neumann boundary conditions''
 \cite{Preetal}.
 We follow this notation.
 That means $\nabla\phi = 0$ perpendicular to the boundary
 $\partial\Omega$ of a domain $\Omega$ where $\phi$ is defined as the
 solution of a partial differential equation $\D \phi = 0$.
 Derivatives are approximated on the lattice by difference quotients.
 The lattice transcription of ``$\nabla\phi = 0$ on $\partial\Omega$''
 is done in such a way that derivative terms $a^{-1}[\phi (z) - \phi
 (z')]$ are omitted in the discretized version of the differential
 operator $\D$ when one site $z\in\Omega$ and the other
 $z'\not\in\Omega$.

 The gauge covariant lattice Laplacian $\LAPN$ with Neumann \bc
 on the boundary of a domain $\Omega$ which equals
 a block~$x$, depends on the lattice gauge field $U$ and is defined by
 \begin{equation}
  (\LAPN \phi ) (z) =
                  \sum_{\stackrel{{\scriptstyle z'\ \subs{\rm n.n.}\ z}}
                                 {z' \in x}}
                    \left[ U(z,z') \phi(z') - \phi (z) \right]
                      \quad \mbox{for $z \in x$.}
  \label{LAPN}
 \end{equation}
 This definition of $\LAPN$ agrees with the one used by Ba{\l}aban
 \cite{BalProp} in rigorous works on constructive gauge theories.
 Summation on the \rhs of Eq.~\equ{LAPN} is over next neighbors $z'$ of
 $z$ which lie in the same block~$x$.

 The imposition of (homogeneous) Dirichlet \bc instead
 of Neumann \bc would result in the lattice operator
 \begin{equation}
  (\LAPD \phi ) (z) =  -2d\,\phi (z) +
                  \sum_{\stackrel{{\scriptstyle z'\ \subs{\rm n.n.}\ z}}
                                 {z' \in x}}
                                 U(z,z') \phi(z')
                      \quad \mbox{for $z \in x$.}
  \label{LAPD}
 \end{equation}
 Note that for a blocking factor $L_b = 2$ the operators $\LAPN$
 and $\LAPD$ differ by a multiple of the unit operator, namely
 $d\,\bbbone$\@.
 In this case a solution of Eq.~\equ{EVequationC} is also a solution of
 the corresponding equation where $\LAPN$ is replaced by $\LAPD$, but
 this is not true when $L_b \neq 2$.

 \smallskip
 The reason why ``ground-state projection MG'' is not an a priori
 defined scheme requires further discussion:
 Ground-state projection MG is supposed to preserve block-locally
 the lowest mode of~$D$.
 In case of bosons, one has to impose \bc for the
 covariant Laplace operator on the boundaries of blocks~$x$.
 This gives us what we called ``a block-local Hamiltonian''.
 However, a priori there exists no distinguished choice of boundary
 conditions so that the meaning of ``block-local Hamiltonian'' is not
 unique.
 Therefore ``ground-state projection MG'' is a priori no completely
 defined scheme.
 The choice of proper boundary conditions is even more involved
 in case of staggered fermions.
 For this discussion we refer to Sec.~\ref{SecBSforSF}.

 In case of bosons we decided to impose Neumann \bc for the following
 reasons.
 We noted already above that Neumann \bc preserve the invariance of the
 free action $\H_0\ =\ < \phi , -\Delta \phi >$ under shifts by constant
 fields when no gauge fields are present.
 In the absence of gauge fields, solutions of the eigenvalue equation
 \equ{EVequationC} are constants on blocks if the subscript $N$ stands
 for Neumann \bc as defined in Eq.~\equ{LAPN}.
 These solutions are known to be ``good'' when one performs
 renormalization group calculations \cite{GawKup,GawKupLH}.
 In the practical method of variational MG for propagators,
 ``piecewise-constant'' kernels are successful in eliminating CSD
 (Sec.~\ref{SecFreePropagators}).
 It is reasonable to start from such requirements on the averaging
 kernel~$C$.
 If one used Dirichlet instead of Neumann \bc in \equ{EVequationC},
 variational MG would not be successful in the limiting case of
 vanishing gauge coupling.

 In order to be precise, in what follows one should always read
 ``ground-state projection MG with Neumann \bc on the boundaries of
 blocks'' instead of just ``ground-state projection MG''.

 \SubSubsection{Averaging Map $C$ for Bosons in Non-Abelian Gauge
                Fields}
 Eq.~\equ{EVequationC} is covariant when $\LAPN$ is the operator
 \equ{LAPN}, because if $\Cstar (z,x)$ is a solution in a given gauge
 field $U$, then $g(z)\,\Cstar (z,x)$ will satisfy \equ{EVequationC} in
 the gauge transformed configuration \equ{GTU}.
 This is so because the eigenvalues of $\LAPN$ are gauge invariant.

 We noted earlier that $C(x,z)$ is an $\Nc \times \Nc$ matrix
 in gauge theories.
 Hence one might say that the eigenvalue equation \equ{EVequationC}
 determines an ``eigenmatrix'' rather than a (1-column) eigenvector.
 In case of gauge group $SU(2)$, Eq.~\equ{EVequationC} has two
 degenerate solutions for any gauge field, when it is regarded as
 an equation for a 1-column vector in place of a matrix.
 To see this suppose that
 $(c_{11} , c_{12} , c_{21} , c_{22} ,\ldots , c_{V1} , c_{V2})^T$,
 $V=L_b^d$\,, is an eigenvector of $-\LAPN$ corresponding to an
 eigenvalue $\lambda$.
 By using the fact that the elements of (a multiple of) an $SU(2)$
 matrix $(U_{ij})_{i,j=1,2}$ fulfill $U_{21} = -\ol{U_{12}}$,
 $U_{22} = \ol{U_{11}}$, one proves that
 $(-\ol{c_{12}} , \ol{c_{11}} , -\ol{c_{22}} , \ol{c_{21}} ,
  \ldots , -\ol{c_{V2}} , \ol{c_{V1}})^T$
 is also an eigenvector of $-\LAPN$ with the same eigenvalue $\lambda$.
 The two independent 2-component solutions may be combined into
 a 2-column matrix $C(x,z)$.
 The freedom of taking linear combinations reflects itself in the
 freedom of performing dielectric gauge transformations on the block
 lattice, i.\ e.\ taking
 \begin{equation}
   \Cstar (z,x) \rightarrow \Cstar (z,x)\,\Upsilon (x) \esk
 \label{Arbitrariness}
 \end{equation}
 where $\Upsilon (x)$ is an arbitrary $2 \times 2$ matrix.
 The freedom of performing dielectric gauge transformations (i.\ e.\
 transformations with matrices which are not elements of the gauge
 group) is eliminated by the normalization condition
 \equ{normalizationC}.
 But this leaves the freedom of performing ordinary (unitary) gauge
 transformations on the block lattice.
 This remaining arbitrariness is eliminated by imposing
 the covariance condition
 \begin{equation}
  C( x , \hat{x} ) = r(x)\,\bbbone \esk \quad \mbox{with $r(x) > 0$
  real.}
  \label{CovarianceCondition}
 \end{equation}
 For gauge groups different from $U(1)$ and $SU(2)$, the \rhs
 of \equ{CovarianceCondition} is to be replaced by a positive Hermitean
 matrix.

 The condition \equ{CovarianceCondition} ensures the transformation
 property \equ{CovKernels}.
 In particular, if $U$ is a pure gauge, then $(L_b a)^d C(x,z)$
 equals the parallel transporter along an arbitrary path from $z$
 to $\hat{x}$.
 $C(x,z)$ is therefore known beforehand in trivial gauge fields.

 \bigskip
 \Subsubsection{Concluding remarks}
 The first use of a ground-state projection definition of the
 restriction operator $C$ was reported by Hulsebos, Smit and Vink
 \cite{HSVGSP} in MG Monte Carlo simulations of a two-dimensional scalar
 $U(1)$ model, and by Brower, Rebbi and Vicari \cite{BRVGSP} for
 propagators in two-dimensional $U(1)$ gauge fields.
 The present author pointed out that the ground-state projection
 method is applicable in arbitrary dimension $d$ and for arbitrary
 gauge group \cite{KalNP2}.
 In Ref.~\cite{KalNP2} an efficient algorithm for solving Eq.
 \equ{EVequationC} was described,
 and it was also proposed to use ground-state projecting kernels in
 Monte Carlo renormalization group studies.
 The lowest eigenvalue $\lambdanull$ of $-\LAPN$ (or $-\LAPD$) is a
 measure of disorder and its renormalization group flow is therefore
 instructive.

 The numerical algorithm for solving Eq. \equ{EVequationC} will
 be explained in Sec.~\ref{SecAlgorithmC}.
 Computations of propagators will be reported in
 Secs.~\ref{SecFreePropagators}--\ref{SecFermionicPropagators}.

 \Subsection{Updating on a Multigrid Layer Consisting of a Single Site
 \label{SecLP}}
 \markboth{{\rm\thesection.}\ {\sc Deterministic Multigrid Methods}}
          {{\rm\thesubsection.}\ {\sl Updating on a Single-Site MG
                                      Layer}}
 {\sl Updating on an MG layer consisting of a single site will be
      investigated.
      We will argue that such an updating can be viewed as a global
      rescaling of an approximate solution $\phin$ of
      Eq.~\equ{propagator} by an $\Nc\times\Nc$ matrix $\Omega$:
      $\phin \mapsto \phin\,\Omega$.
      The matrix $\Omega$ is fixed by the requirement that the energy
      functional of the rescaled propagator gets minimized.
      The gauge covariance properties of this proposal will be discussed
      in detail.}
 \smallskip

 Recall that the CSD of conventional relaxation algorithms for solving
 Eq.~\equ{propagator} depends only on $m^2$ and not on the lattice size,
 Eq.~\equ{scalingtau}.
 Therefore one continues to have CSD on a lattice of only $2^d$ sites,
 and it seems necessary to go to a $1^d$~lattice (a single site) in
 order to eliminate the appearance of CSD\@.
 Of course, a $2^d \Nc \times 2^d \Nc$ matrix can be inverted with
 reasonable effort, but this is no objection against the investigation
 of ``updating on a last single site''.

 The reader should have the bosonic case or Wilson fermions in mind
 when reading this section.
 The treatment of the case of staggered fermions requires some notations
 which have not been introduced yet.
 The discussion for staggered fermions will be deferred to
 Sec.~\ref{SecLPfermi}.

 \Subsubsection{General framework}
 When we update on a $1^d$ sublattice $\LN = \{ x_N \}$,
 interpolation leads to the replacement
  \begin{equation}
    \phin (z) \mapsto \phin (z)
                    + \A (z,x_N)\,\tilde{\Omega} ( x_N) \esp
  \label{lastpoint}
  \end{equation}
 Here we adopt the unigrid point of view, i.~e.\ we consider the
 effect of the MG update on the field on $\Lnull$\@.
 $\A$ denotes a kernel which interpolates directly from a
 $1^d$ sublattice $\LN$ to $\Lnull$\@.
 $\tilde{\Omega} ( x_N)$ is the error of $\phin$ represented at
 the last site $x_N$.
 Note that $\tilde{\Omega} (x_N)$ is an $\Nc\times\Nc$ matrix in
 gauge theories.

 \Subsubsection{Gauge covariance}
 Let us discuss the covariance properties of \equ{lastpoint}, and let
 us assume that the \rhs $f$ of Eq.~\equ{propagator} transforms like
 a matter field \equ{GTphi} under gauge transformations on $\Lnull$.
 For instance, this will be the case in applications with the
 hybrid Monte Carlo algorithm \cite{HybridMC} in simulations of QCD
 with dynamical fermions where $f$ is a pseudofermion field.
 [In the hybrid Monte Carlo algorithm pseudofermion fields $f$ are
 generated by a Gaussian heatbath method.
 One chooses a field $\eta$ from a Gaussian probability
 distribution, and one sets $f = (-\Dirac + m)\,\eta$.
 Since the Gaussian distribution of $\eta$ is invariant under gauge
 transformations $\eta \mapsto g\,\eta$, it follows that pseudofermions
 fields transform according to \equ{GTphi} under gauge transformations.]

 When $f$ transforms like a matter field, the same will be true for
 the smeared propagator $\phi$ and for all the approximate solutions
 $\phin$.
 This is so because iterative algorithms are gauge covariant in the
 sense that all $\phin$ are gauge transformed by $g$ if $g$ is applied
 before relaxation is started.
 (We always begin with an initial $\phi^{(0)} = 0$.)

 We required the interpolation kernel $\A$ to transform under gauge
 transformations on $\Lnull$ according to \equ{CovKernels}.
 However, in case that one considers a layer consisting of a single
 site, the choice of a ``block center $\hat{x}_N$'' is completely
 arbitrary because there is no distinguished site.
 Furthermore, we would like to have the possibility of performing
 updates of the form \equ{lastpoint} also in cases where the linear
 extension of the lattice $\Lnull$ is not a power of~$\L$.
 For example, when $\L = 3$, an $18^4$ lattice can be blocked to
 a $6^4$ lattice and that to a $2^4$ lattice; but one cannot block
 any further if one insists in retaining $\L = 3$.
 Of course, it is not compulsory in an MG algorithm to have the same
 scale factor $\L$ in every blocking step, but in case of staggered
 fermions one is restricted to odd~$\L$.%
 \footnote{In case of staggered fermions the lattice in the above
           example is better replaced by a lattice where one ends up
           with a $4^4$ lattice.
           The reason for this will become clear later on.}
 That follows from considerations concerning the symmetry of free
 staggered fermions and will be explained in Sec.~\ref{SecBSforSF}.

 Actually, one does not have to identify a ``last site $x_N$''
 with a site $\hat{x}_N\in\Lnull$.
 This observation is similar to the point which Palma encountered
 in his computation of constraint effective potentials in gauge
 theories \cite{Pal}.
 Palma pointed out that the constraint effective potential is not
 affected by gauge transformations on a last $1^d$ layer.
 An argument $x_N$ is redundant, and therefore we can follow Palma and
 write $\A (z)$ instead of $\A (z,x_N)$ in \equ{lastpoint}.
 When we consider updates of the form \equ{lastpoint} it is therefore
 legitimate to relax \equ{CovKernels} by requiring that just
 \begin{equation}
      \A (z)  \mapsto  g(z)\,\A (z)
 \label{CovKernelLP}
 \end{equation}
 under gauge transformations $g$ on $\Lnull$.
 Then $\tilde{\Omega}$ has no argument $x_N$ and it must be gauge
 invariant, instead of transforming like a matter field sitting
 at~$\hat{x}_N$.

 \Subsubsection{Choice of the interpolation kernel $\A$}
 Given an approximate solution $\phin$ of Eq.~\equ{propagator}, the
 equation for the error $\en = \phi - \phin$ is $D \en = \rn$, where
 the residual equals $\rn = f - D \phin$.
 We wish to have
  \begin{equation}
   (D \A )(z)\,\tilde{\Omega} = \rn (z) \esk
  \label{lastpointResidual}
  \end{equation}
 but we cannot satisfy this equation, otherwise $\A\tilde{\Omega}$
 would be the error and we were done.
 Therefore we ask ourselves how Eq.~\equ{lastpointResidual} can be
 solved ``as well as possible''.

 Let us introduce the notation
 \begin{equation}
    \bpl \phin\,,\,f \bpr\ \equiv\ \frac{1}{| \Lnull |}
     \sum_{z \in \Lnull} \phin (z)^{\dagger}\,f (z) \esp
 \label{bplbpr}
 \end{equation}
 Note that \equ{bplbpr} defines an $\Nc\times\Nc$ matrix in gauge
 theories.
 (Mind also the dagger~$^{\dagger}$.)

 By multiplying Eq.~\equ{lastpointResidual} from the left with the
 adjoint of some lattice field $\psi$ which will be specified
 below, and summing the result over $z$, we obtain
  \begin{eqnarray}
   \tilde{\Omega}
   & = & \bpl \psi , D \A \bpr^{-1} \cdot \bpl \psi , \rn \bpr
   \nonumber\\
   & = & \bpl \psi , D \A \bpr^{-1} \cdot \bpl \psi , f \bpr
   - \bpl \psi , D \A \bpr^{-1} \cdot \bpl \psi , D \phin\bpr \esp
  \label{TildeOmega}
  \end{eqnarray}
 In the rest of this section and in Sec.~\ref{SecRelatedImpr}
 the dot ``\,$\cdot$\,'' shall emphasize that
 $\Nc\times\Nc$ matrices are multiplied.
 Inserting \equ{TildeOmega} into Eq.~\equ{lastpointResidual} yields
  \begin{equation}
   \int_{z'} D (z,z' ) \cdot \left[
     \A (z') \cdot \bpl \psi , D \A \bpr^{-1} \cdot \bpl \psi , f
             \bpr
 - \A (z') \cdot \bpl\psi , D \A \bpr^{-1} \cdot \bpl \psi , D
           \phin\bpr
     \right] = \rn (z) \esp
  \label{lastpointResidualRewritten}
  \end{equation}
 Compare Eq.~\equ{lastpointResidualRewritten} with the residual
 equation which reads $\int_{z'} D (z,z' ) \cdot \left[ \phi (z' )
 - \phin (z' ) \right] = \rn (z)$.
 The second term in the bracket of \equ{lastpointResidualRewritten}
 equals $\phin$ if we set $\A = \phin$, irrespective of $\psi$.
 This is a legitimate choice which has the correct gauge covariance
 property discussed above.

 By choosing $\A = \phin$, the first term in the bracket of
 \equ{lastpointResidualRewritten} becomes $\phin (z') \cdot \bpl
 \psi , D \phin\bpr^{-1} \cdot \bpl \psi , f \bpr$.
 This term should equal $\phi (z' )$ ``as well as possible''.
 Hence the problem has been reduced to the question how $\psi$
 should be chosen in order to approximate the residual equation
 in an optimal fashion.

 We adopt the variational point of view and take $\phin (z') \cdot \bpl
 \psi , D \phin\bpr^{-1} \cdot \bpl \psi , f \bpr$ with that
 $\psi$ as the best approximation to $\phi (z' )$ where the energy
 functional \equ{EnergyFunctional} is lowest.
 For gauge groups $U(1)$ and $SU(2)$ the computation of this optimal
 $\psi$ is straightforward.
 The result is $\psi = \phin$.

 Note that with $\psi = \phin = \A$, the error $\tilde{\Omega}$ of
 $\phin$ represented at the last site can be written as
 $\tilde{\Omega} = (C D \A )^{-1} \cdot (C \rn )$ with $C = \Astar$.
 This is exactly the same equation which one obtains for
 $\tilde{\Omega}$ when one adopts a variational MG algorithm with
 $C = \Astar$; cf.\ Eqs.~\equ{residualEQcoarse}
 and~\equ{CoarseGridOperator}.

 \Subsubsection{Re-interpretation of Eq.~\equ{lastpoint}}
 By putting everything together, we end up with the following.
 We noted that the argument $x_N$ in \equ{lastpoint}  is redundant, and
 we chose $\A (z) = \phin (z)$.
 Then \equ{lastpoint} reads $\phin (z) \mapsto \phin (z) \cdot
 ( \tilde{\Omega} + \bbbone )$.
 For $\tilde{\Omega}$ we made the choice $\tilde{\Omega} =
 \bpl \phin , D \phin \bpr^{-1} \cdot \bpl \phin , \rn \bpr$.
 Since $\rn = f - D \phin$, one has $\tilde{\Omega} + \bbbone =
 \bpl \phin , D \phin \bpr^{-1} \cdot \bpl \phin , f \bpr
 \equiv \Omega$.

 Summarizing, we re-interpret the updating \equ{lastpoint} as a global
 rescaling of $\phin$ with a matrix $\Omega$ from the right:
  \begin{equation}
    \phin (z) \mapsto \phin (z)\cdot\Omega \quad\mbox{with}\quad
     \Omega = \bpl \phin , D \phin \bpr^{-1}\cdot\bpl \phin , f\bpr \esp
  \label{rescaleBOSE}
  \end{equation}
 We note that the global rescaling by multiplying the vector $\phin$
 with a matrix $\Omega$ from the right is defined because $\phin (z)$
 is an $\Nc\times\Nc$ matrix.
 We also note that the matrix $\Omega$ depends on the iteration
 number $^{(n)}$ but we refrain from indicating that in the notation.

 Since we assumed that $f$ transforms like a matter field under gauge
 transformations, the $\Nc\times\Nc$  matrix $\Omega$ is gauge
 invariant, and \equ{rescaleBOSE} is therefore gauge covariant.

 \smallskip
 In case of gauge group $SU(2)$ the matrix $\bpl \phin , D \phin \bpr$
 is always a multiple of the unit matrix because $D$ is Hermitean.
 In other cases, e.~g.\ in case of gauge group $SU( 3 )$, the matrix
 $\bpl \phin , D \phin \bpr$ can be inverted easily.

 \Subsubsection{The case that $f$ is a $\delta$-function}
 In QCD simulations with the hybrid Monte Carlo algorithm and in
 quenched computations one will not only need the case that $f$ behaves
 like a matter field under gauge transformations, but also the case
 that $f$ is a $\delta$-function.
 Let us discuss the case that $f(z) = \delta_{z,w}\,\bbbone$.
 This $f$ is gauge invariant.

 Then all approximate solutions $\phin$ in iterative algorithms do not
 transform according to \equ{GTphi}, but
 according to $\phin (z) \mapsto g(z)\cdot\phin (z)\cdot g(w)^{-1}$.
 Therefore we cannot proceed as above.
 We are forced to identify $w$ with the ``last site'' when
 we choose $\A = \phin$, but this does not harm.
 The kernel $\A = \phin$ has the gauge covariance property
 \equ{CovKernels} in this case.
 [We remark that here $w$ is a distinguished site which was not
  present in the situation discussed above where $f$ was assumed to
  transform like a matter field.]

 The expression given for $\Omega$ in \equ{rescaleBOSE} will not
 be gauge invariant.
 Rather $\Omega$ will transform like a matter field sitting at site $w$
 in the adjoint representation, i.~e.\ $\Omega \mapsto
 g(w)\cdot\Omega\cdot g(w)^{-1}$.
 However, this is exactly what is required to make the operation
 $\phin \mapsto \phin\cdot\Omega$ gauge covariant.
 So everything is o.~k.\ also in the case that $f$ is a
 $\delta$-function.

 \Subsubsection{Outlook to the case of staggered fermions}
 It was noted above that \equ{rescaleBOSE} will have to be modified
 in case of staggered fermions.
 This modification will be discussed in Sec.~\ref{SecLPfermi}.
 Here we remark that gauge covariance of the fermionic analog of
 \equ{rescaleBOSE} is ensured in applications with the
 hybrid Monte Carlo algorithm \cite{HybridMC}.

 Suppose that the \rhs $f$ in Eq.~\equ{propagator} is a pseudofermion
 field.
 The fermionic matrices $\Omega$ are given by similar expressions as in
 \equ{rescaleBOSE}.
 These expressions are also gauge invariant.
 Therefore a global rescaling $\phin \mapsto \phin\cdot\Omega$ is also
 gauge covariant in case of staggered fermions.
 When $f$ is a $\delta$-function, analogous remarks as above apply.

 \medskip
 \Subsubsection{Concluding remarks}
 Note that $\Omega\rightarrow\bbbone$ and that the error represented
 at the last site $\tilde{\Omega}\rightarrow 0$, as $\phin\rightarrow
 \phi$.
 A finding of numerical experiments reported later is that in practice
 $\Omega = \bbbone$ as soon as errors decay exponentially.
 Then the step \equ{rescaleBOSE} can be switched off.

 The expression given for the matrix $\Omega$ in \equ{rescaleBOSE}
 is the same which one obtains when one starts from replacing $\phin$
 by $(\phin\cdot\Omega )$, with $\Omega$ chosen such that the energy
 functional \equ{EnergyFunctional} of the rescaled approximation
 $\phin\cdot\Omega$ gets minimized.

 The proposal \equ{rescaleBOSE} was made in a recent paper
 \cite{KalLastPoint}.
 Numerical results of its implementation will be presented in
 Secs.~\ref{SecFreePropagators7}, \ref{SecLPbose}, and~\ref{SecLPfermi}.

 We add as a comment that inclusion of \equ{rescaleBOSE} in a CG
 iteration would have no effect.
 The CG iterates $\phin$ have the property that $K[\phin\cdot\Omega ]$
 is minimal for $\Omega = \bbbone$.

 \Subsection{Related Improvements from a Variational Point of View
 \label{SecRelatedImpr}}
 \markboth{{\rm\thesection.}\ {\sc Deterministic Multigrid Methods}}
          {{\rm\thesubsection.}\ {\sl Related Improvements from a
                                      Variational Point of View}}
 {\sl Some gauge covariant modifications of (MG) relaxation algorithms
      which are related to the updating on an MG layer consisting of a
      single site as described in Sec.~\ref{SecLP} will be discussed
      now.
      The new parameters are not tunable, they are all determined
      by the algorithms themselves.
      The modifications require additional computational work, though.}
 \smallskip

 \Subsubsection{Modified MG correction updating step}
 In conventional MG approaches one considers updates of the form
 $\phin \mapsto \phin + \vpn$ where $\vpn$ is obtained by interpolation
 of an approximate solution of a residual equation on a coarser lattice.
 We propose to generalize this to
  \begin{equation}
    \phin (z) \mapsto \phint \equiv
     \phin (z)\cdot\Omega + \vpn (z)\cdot\Theta \esp
  \label{ImprovedCorr}
  \end{equation}
 The two $\Nc \times \Nc$ matrices $\Omega$ and $\Theta$ are
 chosen such that the energy functional $K [ \phint ]$ is minimized.
 In particular, this proposal may be an improvement in algorithms
 where the residual equation is only solved approximately,
 or in algorithms were coarse grid operators are not defined through
 the Galerkin prescription.

 \Subsubsection{Modified checkerboard SOR}
 Consider for illustration the bosonic problem.
 When we update at the even sites, we propose to modify checkerboard
 SOR \equ{checkerboardSOR} according to
  \begin{eqnarray}
    \phin (z)   \mapsto   \phin (z)\cdot\Omega & +\ \ \vpn (z)\cdot
                       \Theta & \mbox{if $z$ is even} \esk \nonumber \\
    & & \label{SOR} \\
    \phin (z)   \mapsto   \phin (z)\cdot\Xi    &                       &
              \mbox{if $z$ is odd} \esk  \nonumber
  \end{eqnarray}
 where $\vpn (z) = ( 2d + m^2)^{-1} [ f(z) + \sum_{z'\mbox{\scriptsize
 n.n.}z} U ( z , z' ) \cdot \phin ( z' ) ]$.
 Again, the three $\Nc\times\Nc$ matrices $\Omega$, $\Theta$ and $\Xi$
 should be chosen such that the functional $K$ gets minimized.
 The proposal \equ{SOR} expresses the view that in gauge theories one
 should have gauge invariant/covariant relaxation matrices rather
 than relaxation parameters that are real numbers.

 \Subsubsection{Modified damped Jacobi relaxation}
 Damped Jacobi relaxation can be generalized according to
 \equ{ImprovedCorr} with $\vpn = (2d + m^2)^{-1}\,\rn$.
 If one fixes $\Omega = \bbbone$, one recovers (in case of $U(1)$ and
 $SU(2)$ gauge fields) the minimal residual algorithm that was used by
 Hulsebos et al.\ \cite{HSVLAT90,HSVNP,HSVJulich,Vin,VinLAT91}.

 \smallskip
 \Subsubsection{Gauge covariance}
 We remark that also the proposals
 \equ{ImprovedCorr} and \equ{SOR} respect gauge covariance.
 The matrices introduced here are gauge invariant when $f$ transforms
 like a matter field under gauge transformations (e.~g.\ when $f$
 is a pseudofermion field).
 When $f$ is a $\delta$-function, the matrices transform like a matter
 field in the adjoint representation, as discussed in
 Sec.~\ref{SecLP}.

 \smallskip
 {}From now on we will refrain again from explicitly indicating the
 multiplication of matrices by a dot.

 \Subsectiontocentry{Cautionary Remarks Concerning the Term ``Critical
                     Slowing Down''}
                     {Cautionary Remarks Concerning the Term \protect\\
                      ``Critical Slowing Down''
 \label{SecCautionaryRemark}}
 \markboth{{\rm\thesection.}\ {\sc Deterministic Multigrid Methods}}
          {{\rm\thesubsection.}\ {\sl Cautionary Remarks Concerning the
                                      Term CSD}}
 {\sl It will be pointed out that for some algorithms the asymptotic
      CSD behavior can be predicted from studies on lattices of fixed
      size with small values of $\Dm$.}
 \smallskip

 Numerical results which will be presented in subsequent sections
 show that in case of traditional relaxation algorithms for bosonic
 propagators the predicted CSD behavior of~$\tau$, discussed in
 Sec.~\ref{SecCSD1grid}, can be monitored extremely well in practice.
 An important point is that the constant of proportionality in the
 scaling relation \equ{scalingtau} does practically {\em not} depend on
 the lattice size; e.~g.\ in Jacobi relaxation it equals
 $(2d + \mcr )$\,,Eq.~\equ{scalingJac}.
 The quantity $\mcr$ depends only slightly on the lattice size, and
 smoothly on $\beta$, but $(2d + \mcr)$ depends on $| \Lambda |$, or
 on~$\beta$, only very weakly.
 (We are interested in $d=4$.)

 Therefore the study of the {\em asymptotic} behavior of $\tau$
 under conditions where the linear extension of the lattice and
 $1/\triangle m$ are changed proportionally (scaling conditions), can
 be predicted from studies in a fixed volume.
 Certainly, in physical applications (Monte Carlo simulations) the
 inverse mass should be smaller than the extension of the lattice,
 but this is an aspect of finite size effects on physical observables.
 The existence of the $1 / \Dm$ divergence (with a constant independent
 of the lattice size) on a lattice of fixed size implies the appearance
 of CSD in computations where all quantities are scaled appropriately.

 \smallskip
 To the author's knowledge there exists no study in the literature
 where $\mcr$ is not disregarded in case of staggered fermions.
 This neglect is only justified by the smallness of $\mcr$ but it has
 never been checked whether the neglect is justified.
 A result of the present work is the validity of the relation $\tau =
 \mbox{const}/\Dm$ in traditional one-grid and variational MG
 relaxation,%
 \footnote{In order to be precise, one should say variational MG with
           the ``Laplacian choice'' of the averaging operator~$C$.
           This choice will be introduced in Sec.~\ref{SecBSforSF}.}
 with a constant which is also {\em independent of the lattice size\/}.
 Therefore also for fermions the CSD of $\tau$ in these algorithms
 under scaling conditions can be predicted equivalently from studies on
 small lattices.

 By neglecting $\mcr$ and trying to determine $z$ under scaling
 conditions on small lattices ($\lsim 18^4$), one can at most obtain
 some effective critical exponent $z_{\subs{eff}}$.
 This $z_{\subs{eff}}$ contains however a great deal of arbitrariness
 and cannot be defined uniquely.
 One can run into difficulties with this procedure~\cite{HSVNP}.
 The author admits that a $z_{\subs{eff}}$ is of more practical
 relevance as long as numerical simulations are limited to
 lattice sizes where $m$ is not really small.
 But we look for algorithms which can be used in future large scale
 computations, and for these it will be $z$ and not $z_{\subs{eff}}$
 which governs CSD\@.

 There is one weak point in this reasoning.
 It is the disregard of a possibly remaining volume effect with respect
 to how many iterations it takes until decays of errors are
 exponential.
 This volume effect has to be taken into consideration
 when one wants to estimate CPU times.
 The volume effect also shows up in CG where it must not be disregarded,
 otherwise CG would be an ideal algorithm because in practice its
 ``asymptotic'' convergence rate is very rapid.
 Therefore one might have to go back to a~$z_{\subs{eff}}$\/.
 We will see later that such a volume effect is only appreciable when
 the proposal \equ{rescaleBOSE} and its generalization for staggered
 fermions is included in algorithms.
 For this discussion we refer to Secs.~\ref{SecLPbose}
 and~\ref{SecLPfermi}.

 \Section{Gauge Covariant Block Spins
 \label{SecBS}}
 \markboth{{\rm\thesection.}\ {\sc Gauge Covariant Block Spins}}{}
 {\sl In this section we will discuss gauge covariant block averages
      for bosons, for staggered fermions and for gauge fields.
      In case of staggered fermions the blocking procedure is not
      obvious.
      If one wants to preserve as much as possible of the lattice
      symmetry group of free staggered fermions, one is, in particular,
      forced to choose an odd scale factor $L_b$\,; a factor-of-2
      coarsening is therefore not allowed.
      An efficient numerical algorithm for the computation of the
      ground-state projecting averaging kernels will be presented.
      After that, results are given for optimal interpolation kernels in
      $SU(2)$ gauge fields in four dimensions.
      They prove that ground-state projection is a good choice of
      block spin in the presence of gauge fields.
      As a supplement a short review of Ba{\l}aban's analytical work on
      decay of propagators in gauge fields is given.
      Finally the notion of ``smoothness'' in gauge theories is
      discussed.}

 \Subsection{Block Spin for Bosons
 \label{SecBSforBosons}}
 \markboth{{\rm\thesection.}\ {\sc Gauge Covariant Block Spins}}
          {{\rm\thesubsection.}\ {\sl Block Spin for Bosons}}
 {\sl Block spins for bosons are defined by means of the ground-state
      projecting kernel~$C$ described in Sec.~\ref{SecGSP}.
      Probability distributions of $\|C(x,z)\|$ in four-dimensional
      $SU(2)$ gauge fields are given.}
 \smallskip

 For bosonic matter fields $\phi$ one can define block lattices in
 the ``naive'' straightforward way which was described in
 Sec.~\ref{SecMGalgorithms}.
 This blocking is illustrated in Fig.~\NextFig\
 for a two-dimensional lattice.
 \begin{figure}
 \setlength{\unitlength}{1.0mm}
 \begin{minipage}{16.5 cm}{}
 \begin{center}
 \begin{picture}(80,80)(2.0,2.0)
 \multiput(10,10)(8,0){9}{\Huge$\bullet$}
 \multiput(10,18)(8,0){9}{\Huge$\bullet$}
 \multiput(10,26)(8,0){9}{\Huge$\bullet$}
 \multiput(10,34)(8,0){9}{\Huge$\bullet$}
 \multiput(10,42)(8,0){9}{\Huge$\bullet$}
 \multiput(10,50)(8,0){9}{\Huge$\bullet$}
 \multiput(10,58)(8,0){9}{\Huge$\bullet$}
 \multiput(10,66)(8,0){9}{\Huge$\bullet$}
 \multiput(10,74)(8,0){9}{\Huge$\bullet$}
 \thinlines
 \multiput(19.75,19.95)(24,0){3}{\circle{6}}
 \multiput(19.75,43.95)(24,0){3}{\circle{6}}
 \multiput(19.75,67.95)(24,0){3}{\circle{6}}
 \thicklines
 \put(8.5,8.5){\dashbox{0.7}(22,22){\mbox}}
 \put(32.5,8.5){\dashbox{0.7}(22,22){\mbox}}
 \put(56.5,8.5){\dashbox{0.7}(22,22){\mbox}}
 \put(8.5,32.5){\dashbox{0.7}(22,22){\mbox}}
 \put(32.5,32.5){\dashbox{0.7}(22,22){\mbox}}
 \put(56.5,32.5){\dashbox{0.7}(22,22){\mbox}}
 \put(8.5,56.5){\dashbox{0.7}(22,22){\mbox}}
 \put(32.5,56.5){\dashbox{0.7}(22,22){\mbox}}
 \put(56.5,56.5){\dashbox{0.7}(22,22){\mbox}}
  \thinlines\put(3.7,11.7){\line(1,0){8}}
 \thicklines\put(11.7,11.7){\line(1,0){16}}
  \thinlines\put(27.7,11.7){\line(1,0){8}}
 \thicklines\put(35.7,11.7){\line(1,0){16}}
  \thinlines\put(51.7,11.7){\line(1,0){8}}
 \thicklines\put(59.7,11.7){\line(1,0){16}}
  \thinlines\put(75.7,11.7){\line(1,0){8}}
  \thinlines\put(3.7,19.7){\line(1,0){8}}
 \thicklines\put(11.7,19.7){\line(1,0){16}}
  \thinlines\put(27.7,19.7){\line(1,0){8}}
 \thicklines\put(35.7,19.7){\line(1,0){16}}
  \thinlines\put(51.7,19.7){\line(1,0){8}}
 \thicklines\put(59.7,19.7){\line(1,0){16}}
  \thinlines\put(75.7,19.7){\line(1,0){8}}
  \thinlines\put(3.7,27.7){\line(1,0){8}}
 \thicklines\put(11.7,27.7){\line(1,0){16}}
  \thinlines\put(27.7,27.7){\line(1,0){8}}
 \thicklines\put(35.7,27.7){\line(1,0){16}}
  \thinlines\put(51.7,27.7){\line(1,0){8}}
 \thicklines\put(59.7,27.7){\line(1,0){16}}
  \thinlines\put(75.7,27.7){\line(1,0){8}}
  \thinlines\put(3.7,35.7){\line(1,0){8}}
 \thicklines\put(11.7,35.7){\line(1,0){16}}
  \thinlines\put(27.7,35.7){\line(1,0){8}}
 \thicklines\put(35.7,35.7){\line(1,0){16}}
  \thinlines\put(51.7,35.7){\line(1,0){8}}
 \thicklines\put(59.7,35.7){\line(1,0){16}}
  \thinlines\put(75.7,35.7){\line(1,0){8}}
  \thinlines\put(3.7,43.7){\line(1,0){8}}
 \thicklines\put(11.7,43.7){\line(1,0){16}}
  \thinlines\put(27.7,43.7){\line(1,0){8}}
 \thicklines\put(35.7,43.7){\line(1,0){16}}
  \thinlines\put(51.7,43.7){\line(1,0){8}}
 \thicklines\put(59.7,43.7){\line(1,0){16}}
  \thinlines\put(75.7,43.7){\line(1,0){8}}
  \thinlines\put(3.7,51.7){\line(1,0){8}}
 \thicklines\put(11.7,51.7){\line(1,0){16}}
  \thinlines\put(27.7,51.7){\line(1,0){8}}
 \thicklines\put(35.7,51.7){\line(1,0){16}}
  \thinlines\put(51.7,51.7){\line(1,0){8}}
 \thicklines\put(59.7,51.7){\line(1,0){16}}
  \thinlines\put(75.7,51.7){\line(1,0){8}}
  \thinlines\put(3.7,59.7){\line(1,0){8}}
 \thicklines\put(11.7,59.7){\line(1,0){16}}
  \thinlines\put(27.7,59.7){\line(1,0){8}}
 \thicklines\put(35.7,59.7){\line(1,0){16}}
  \thinlines\put(51.7,59.7){\line(1,0){8}}
 \thicklines\put(59.7,59.7){\line(1,0){16}}
  \thinlines\put(75.7,59.7){\line(1,0){8}}
  \thinlines\put(3.7,67.7){\line(1,0){8}}
 \thicklines\put(11.7,67.7){\line(1,0){16}}
  \thinlines\put(27.7,67.7){\line(1,0){8}}
 \thicklines\put(35.7,67.7){\line(1,0){16}}
  \thinlines\put(51.7,67.7){\line(1,0){8}}
 \thicklines\put(59.7,67.7){\line(1,0){16}}
  \thinlines\put(75.7,67.7){\line(1,0){8}}
  \thinlines\put(3.7,75.7){\line(1,0){8}}
 \thicklines\put(11.7,75.7){\line(1,0){16}}
  \thinlines\put(27.7,75.7){\line(1,0){8}}
 \thicklines\put(35.7,75.7){\line(1,0){16}}
  \thinlines\put(51.7,75.7){\line(1,0){8}}
 \thicklines\put(59.7,75.7){\line(1,0){16}}
  \thinlines\put(75.7,75.7){\line(1,0){8}}
  \thinlines\put(11.7,3.7){\line(0,1){8}}
 \thicklines\put(11.7,11.7){\line(0,1){16}}
  \thinlines\put(11.7,27.7){\line(0,1){8}}
 \thicklines\put(11.7,35.7){\line(0,1){16}}
  \thinlines\put(11.7,51.7){\line(0,1){8}}
 \thicklines\put(11.7,59.7){\line(0,1){16}}
  \thinlines\put(11.7,75.7){\line(0,1){8}}
  \thinlines\put(19.7,3.7){\line(0,1){8}}
 \thicklines\put(19.7,11.7){\line(0,1){16}}
  \thinlines\put(19.7,27.7){\line(0,1){8}}
 \thicklines\put(19.7,35.7){\line(0,1){16}}
  \thinlines\put(19.7,51.7){\line(0,1){8}}
 \thicklines\put(19.7,59.7){\line(0,1){16}}
  \thinlines\put(19.7,75.7){\line(0,1){8}}
  \thinlines\put(27.7,3.7){\line(0,1){8}}
 \thicklines\put(27.7,11.7){\line(0,1){16}}
  \thinlines\put(27.7,27.7){\line(0,1){8}}
 \thicklines\put(27.7,35.7){\line(0,1){16}}
  \thinlines\put(27.7,51.7){\line(0,1){8}}
 \thicklines\put(27.7,59.7){\line(0,1){16}}
  \thinlines\put(27.7,75.7){\line(0,1){8}}
  \thinlines\put(35.7,3.7){\line(0,1){8}}
 \thicklines\put(35.7,11.7){\line(0,1){16}}
  \thinlines\put(35.7,27.7){\line(0,1){8}}
 \thicklines\put(35.7,35.7){\line(0,1){16}}
  \thinlines\put(35.7,51.7){\line(0,1){8}}
 \thicklines\put(35.7,59.7){\line(0,1){16}}
  \thinlines\put(35.7,75.7){\line(0,1){8}}
  \thinlines\put(43.7,3.7){\line(0,1){8}}
 \thicklines\put(43.7,11.7){\line(0,1){16}}
  \thinlines\put(43.7,27.7){\line(0,1){8}}
 \thicklines\put(43.7,35.7){\line(0,1){16}}
  \thinlines\put(43.7,51.7){\line(0,1){8}}
 \thicklines\put(43.7,59.7){\line(0,1){16}}
  \thinlines\put(43.7,75.7){\line(0,1){8}}
  \thinlines\put(51.7,3.7){\line(0,1){8}}
 \thicklines\put(51.7,11.7){\line(0,1){16}}
  \thinlines\put(51.7,27.7){\line(0,1){8}}
 \thicklines\put(51.7,35.7){\line(0,1){16}}
  \thinlines\put(51.7,51.7){\line(0,1){8}}
 \thicklines\put(51.7,59.7){\line(0,1){16}}
  \thinlines\put(51.7,75.7){\line(0,1){8}}
  \thinlines\put(59.7,3.7){\line(0,1){8}}
 \thicklines\put(59.7,11.7){\line(0,1){16}}
  \thinlines\put(59.7,27.7){\line(0,1){8}}
 \thicklines\put(59.7,35.7){\line(0,1){16}}
  \thinlines\put(59.7,51.7){\line(0,1){8}}
 \thicklines\put(59.7,59.7){\line(0,1){16}}
  \thinlines\put(59.7,75.7){\line(0,1){8}}
  \thinlines\put(67.7,3.7){\line(0,1){8}}
 \thicklines\put(67.7,11.7){\line(0,1){16}}
  \thinlines\put(67.7,27.7){\line(0,1){8}}
 \thicklines\put(67.7,35.7){\line(0,1){16}}
  \thinlines\put(67.7,51.7){\line(0,1){8}}
 \thicklines\put(67.7,59.7){\line(0,1){16}}
  \thinlines\put(67.7,75.7){\line(0,1){8}}
  \thinlines\put(75.7,3.7){\line(0,1){8}}
 \thicklines\put(75.7,11.7){\line(0,1){16}}
  \thinlines\put(75.7,27.7){\line(0,1){8}}
 \thicklines\put(75.7,35.7){\line(0,1){16}}
  \thinlines\put(75.7,51.7){\line(0,1){8}}
 \thicklines\put(75.7,59.7){\line(0,1){16}}
  \thinlines\put(75.7,75.7){\line(0,1){8}}
 \end{picture}
 \end{center}
 \end{minipage}
 \caption{{\protect\small{\em
          Division of a two-dimensional bosonic lattice $\Lambda$ into
          blocks.
          Sites $z \in \Lambda$ are marked by {\protect\Huge$\bullet$}.
          A scale factor of $L_b = 3$ is chosen.
          Blocks $x$ are bordered by dashed lines and may be
          identified with the block centers $\hat{x}$ which are
          encircled.
          Links inside blocks are drawn with thick lines, while links
          which connect sites in different blocks are drawn with thin
          lines.}}}
 \setcounter{FigBlockBose}{\value{figure}}
 \setcounter{SecBlockBose}{\value{section}}
 \end{figure}
 Block spins are defined as in Eq.~\equ{BlockAverage}, where the
 restriction operator $C$ is defined by the ground-state projection
 definition \equ{EVequationC}, \equ{normalizationC}, and
 \equ{CovarianceCondition}.
 The links, where both endpoints are in the same block and which yield
 contributions to the Laplacian with Neumann boundary conditions
 \equ{LAPN}, are drawn with thick lines in Fig.~\PreviousFig.
 The links drawn with thin lines connect neighboring blocks.
 The gauge field on these links is not relevant for~$C$.

 \Figure{10.0cm}
        {Averaging kernel $C$ for bosons:
         Probability distributions of the dimensionless quantity
         $(L_b a)^d\,\normC$\,, $z\!\in\!x$, in four-dimensional pure
         $SU(2)$ gauge theory.
         The norm equals $\normC = [ \frac{1}{2} \Tr C(x,z)^{\dagger}
         C(x,z) ]^{1/2}$.
         The four curves shown correspond to $\beta = 10.0$, $5.0$,
         $3.0$, $2.7$, with increasing width.
         In a pure gauge ($\beta = \infty$),\ $(L_b a)^d\,\normC \equiv
         1$ for $z\!\in\!x$.}

 In Sec.~\ref{SecOptA} it will be shown that ground-state projection
 defines a ``good'' block spin for bosons.
 This means that the associated optimal interpolation kernel
 $\A$ decays exponentially (see also Appendix~\theappKernels).
 It was mentioned in Sec.~\ref{SecInclusionGF} that $C(x,z)$ will not be
 an element of the gauge group in general.
 Rather, $C(x,z)$ will be in the linear span%
   \footnote{i.\ e.\ $C(x,z)$ will be a real multiple of an element of
             $SU(2)$ if $G = SU(2)$, an arbitrary complex $N \times N$
             matrix for $G = U(N)$, $N \geq 3$ etc;
             see Ref.~\cite{MacDLGT}.}
 of the gauge group.
 This fact is shown in Fig.~$3.2$ for four-dimensional $SU(2)$.
 The probability distributions were obtained from gauge field
 configurations on $18^4$ lattices which were equilibrated with the
 Wilson action \cite{Wil}, $\H (U) = \beta \sum_P [ 1 - \frac{1}{2} \Tr
 U(P) ]$\@.
 Brower, Rebbi and Vicari \cite{BRVvariational} proposed a
 variational method for approximating ground-state projection kernels.
 Their variational principle restricts $(L_b a)^d\,\normC$ to 1.
 For finite $\beta$ the solution of Eq.~\equ{EVequationC} is not well
 approximated by this procedure, but this does not imply that
 the performance of MG algorithms with this choice is necessarily bad.

 \Subsection{Block Spins for Staggered Fermions
 \label{SecBSforSF}}
 \markboth{{\rm\thesection.}\ {\sc Gauge Covariant Block Spins}}
          {{\rm\thesubsection.}\ {\sl Block Spins for Staggered
                                      Fermions}}
 {\sl One of the first questions which has to be answered in MG
      algorithms is the question for the choice of block lattices.
      The answer to this question is obvious in case of bosons, but it
      is not straightforward in case of staggered fermions.
      Different strategies were pursued in the literature
      \cite{BenBraSol,HSVLAT90}.
      The blocking procedure proposed in this section starts from
      the requirement that as much as possible is preserved of internal
      and space-time symmetries in the limiting case of vanishing gauge
      coupling.
      As a result we are forced to choose a scale factor of~$3$
      (or any other odd number).
      Another consequence is the emergence of seemingly overlapping
      blocks.
      In the limiting case of a pure gauge the blocks have actually no
      sites in common.
      But in nontrivial gauge fields the symmetries of free staggered
      fermions are broken, and for this reason one cannot a priori rule
      out the possibility that the blocks overlap in a nontrivial way.
      Two qualitatively different proposals will be made for the choice
      of the restriction operator~$C$.
      It will also be defined as solution of a gauge covariant
      eigenvalue equation.
      We will call the two proposals the ``Laplacian choice'' and the
      ``Diracian choice''.
      Both proposals reduce to a successful construction in the
      limiting case of free staggered fermions.
      The Diracian choice is preferable for physical reasons, but it is
      computationally quite demanding.
      The reason is that blocks overlap in a nontrivial way, i.~e.\
      every fine grid site makes contributions to the block spin
      at more than just one coarse grid site.
      The Laplacian choice is more convenient from the numerical point
      of view.
      It retains the property that also in nontrivial gauge fields
      the seemingly overlapping blocks have actually no sites in
      common, i.~e.\ every fine grid site makes a contribution to the
      average at exactly one coarse grid site.}
 \smallskip

 \noindent
 This section is based on joint work with G.\,Mack and M.\,Speh
 \cite{KalMacSpe}.

 Staggered fermions reduce the number of doublers of the naive
 lattice fermion action while retaining some aspects of chiral
 symmetry \cite{Sus,ShaThuWei,KluMorNapPet}.
 The idea is to reduce the number of spinor/flavor components at each
 lattice site to one.
 For simplicity, let the dimension~$d$ of space-time be even.
 ($d=4$ is the case we are interested in.)
 The different fermionic degrees of freedom are distributed on the
 sites of hypercubes of volume $2^d$ sites.

 In terms of crystallography as used in solid state physics
 \cite{AshMer}, the lattice of staggered fermions is a hypercubic
 lattice $\Lambda_a$ of unit cells of lattice spacing~$a$ with a
 $2^d$-point basis.
 The basis is given by the sites of a hypercube of volume $(a/2)^d$.
 Sites within the basis are distinguished by their ``pseudoflavor''~$H$.
 Different pseudoflavors would correspond to different species of
 atoms or ions in the solid state analog.
 If one does not distinguish different pseudoflavors, one would
 have a hypercubic lattice $\Lah$ of lattice spacing $a/2$.
 \newline
 A summary about staggered fermions can be found in
 Appendix~\theappSF, together with our notations and conventions.

 \SubSubsection{Blocking Consistent with the Symmetries
                of Free Staggered Fermions}
 Pseudoflavor $H$ is specified by an ordered set of distinct indices
 $\mu_i$,
  \begin{equation}
   H = \{ \mu_1 , \ldots , \mu_h\ |\  0 \leq h < d,\ \mu_1 < \cdots
           \mu_h \} \esp
  \label{PF}
  \end{equation}
 Sites $z = z_H \in \Lah$ with pseudoflavor $H$ are of the form
  \begin{equation}
    z_H = \tilde{y} + \txeh e_H \esk \quad
          e_H = \sum_{\mu \in H} e_{\mu} \esk
  \label{zPF}
  \end{equation}
 where $e_{\mu}$ is a lattice vector of length $a$ in
 $\mu$-direction in $\Lambda_a$.
 $\tilde{y}$ is a site in $\Lambda_a$ which is identified with site
 $z_{\emptyset} \in \Lah$.
 The sites $z_H$ with fixed $H$ form a sublattice $\Lambda_a^H$ of
 $\Lah$, viz.\ $\Lambda_a^H = \Lambda_a + \txeh e_H$.
 [In principle any sublattice $\Lambda_a^H$ could be identified with
  $\Lambda_a$.]
 Fig.~\NextFig\ illustrates this.
 \setlength{\unitlength}{1.0mm}
 \begin{figure}
 \begin{minipage}{16.0 cm}{}
 \begin{center}
 \begin{picture}(90,90)(5.75,5.75)
 \multiput(20,20)(40,0){2}{\Huge$\bullet$}
 \multiput(40,20)(40,0){2}{\Huge$\circ$}
 \multiput(20,40)(40,0){2}{\Huge$\diamond$}
 \multiput(40,40)(40,0){2}{\Huge$\ast$}
 \multiput(20,60)(40,0){2}{\Huge$\bullet$}
 \multiput(40,60)(40,0){2}{\Huge$\circ$}
 \multiput(20,80)(40,0){2}{\Huge$\diamond$}
 \multiput(40,80)(40,0){2}{\Huge$\ast$}
 \put(11.7,21.7){\line(1,0){80}}
 \put(11.7,61.7){\line(1,0){80}}
 \put(21.7,11.7){\line(0,1){80}}
 \put(61.7,11.7){\line(0,1){80}}
 \multiput(12.7,41.7)(4,0){20}{\line(1,0){2}}
 \multiput(12.7,81.7)(4,0){20}{\line(1,0){2}}
 \multiput(41.7,12.7)(0,4){20}{\line(0,1){2}}
 \multiput(81.7,12.7)(0,4){20}{\line(0,1){2}}
 \put(26.5,15){\large$\tilde{y}\!=\!z_{\emptyset}$}
 \put(27,17){\vector(-1,1){4.0}}
 \put(46.5,15){\large$\tilde{y}\!+\!\txeh\,e_1$}
 \put(47,17){\vector(-1,1){4.0}}
 \put(26.5,35){\large$\tilde{y}\!+\!\txeh\,e_2$}
 \put(27,37){\vector(-1,1){4.0}}
 \put(46.5,35){\large$\tilde{y}\!\!+\!\!\txeh\,e_{12}$}
 \put(47,37){\vector(-1,1){4.0}}
 \put(95.0,40.7){\large$a$}
 \put(92.0,70.2){\large$a/2$}
 \put(96.0,43.7){\vector(0,1){18.0}}
 \put(96.0,38.7){\vector(0,-1){17.0}}
 \put(96.0,74.7){\vector(0,1){7.0}}
 \put(96.0,68.7){\vector(0,-1){7.0}}
 \end{picture}
 \end{center}
 \end{minipage}
 \caption{{\protect\small{\em
          Distribution of pseudoflavor degrees of freedom in a
          two-dimensional lattice.
          Sites of pseudoflavor sublattices $\Lambda_a^H$, $H =
          \emptyset, \{ 1 \} , \{ 2 \} , \{ 1,2 \}$,  are marked
          with {\protect\Huge$\bullet$}, {\protect\Huge$\circ$},
          {\protect\Huge$\diamond$}, {\protect\Huge$\ast$},
          respectively.}}}
 \setcounter{FigPFlattice}{\value{figure}}
 \setcounter{SecPFlattice}{\value{section}}
 \end{figure}